\documentclass{jfm}
\usepackage{graphicx}
\usepackage{bm}
\usepackage{xcolor,tabu,amsmath}
\usepackage{epstopdf, epsfig}
\usepackage[colorlinks=true,
            linkcolor=blue,
            urlcolor=blue,
            citecolor=blue]{hyperref}

\shorttitle{Self-organized oscillations of Leidenfrost drops}
\shortauthor{X. Ma, and J. C. Burton}

\title{Self-organized oscillations of Leidenfrost drops}

\author{Xiaolei Ma\aff{}
  \corresp{\email{xiaolei.ma@emory.edu}},
  Justin C. Burton\aff{}
  }

\affiliation{\aff{}Department of Physics, Emory University, Atlanta, Georgia 30322, USA
}

\begin{document}

\maketitle

\begin{abstract}
In the Leidenfrost effect, a thin layer of evaporated vapor forms between a liquid and a hot solid. The complex interactions between the solid, liquid, and vapor phases can lead to rich dynamics even in a single Leidenfrost drop. Here we investigate the self-organized oscillations of Leidenfrost drops that are excited by a constant flow of evaporated vapor beneath the drop. We show that for small Leidenfrost drops, the frequency of a recently reported ``breathing mode" (Caswell, \textit{Phys. Rev. E}, vol. 90, 2014, 013014) can be explained by a simple balance of gravitational and surface tension forces. For large Leidenfrost drops, azimuthal star-shaped oscillations are observed. Our previous work showed how the coupling between the rapid evaporated vapor flow and the vapor-liquid interface excites the star oscillations (Ma \textit{et al.,} \textit{Phys. Rev. Fluids}, vol. 2, 2017, 031602). In our experiments, star-shaped oscillation modes of $n=2$ to 13 are observed in different liquids, and the number of observed modes depends sensitively on the viscosity of the liquid. Here we expand on this work by directly comparing the oscillations with theoretical predictions, as well as show how the oscillations are initiated by a parametric forcing mechanism through pressure oscillations in the vapor layer. The pressure oscillations are driven by the capillary waves of a characteristic wavelength beneath the drop. These capillary waves can be generated by a large shear stress at the liquid-vapor interface due to the rapid flow of evaporated vapor. We also explore potential effects of thermal convection in the liquid. Although the measured Rayleigh number is significantly larger than the critical Rayleigh number, the frequency (wavelength) of the oscillations depends only on the capillary length of the liquid, and is independent of the drop radius and substrate temperature. Thus convection seems to play a minor role in Leidenfrost drop oscillations, which are mostly hydrodynamic in origin.
\end{abstract}


\section{Introduction}

When a volatile liquid drop is deposited on a sufficiently hot solid, it can survive for minutes due to the presence of a thermally-insulating layer of evaporated vapor beneath the drop. In this levitated state, commonly known as the Leidenfrost regime \citep{leidenfrost1756aquae}, the supporting vapor layer is maintained by the sustained evaporation of the liquid, and individual drops are free to undergo frictionless motion due to the absence of liquid-solid contact. The Leidenfrost effect can be easily observed by placing a water drop onto a hot pan over a cook stove in the kitchen, and has been the subject of numerous fundamental and applied studies due to the complex and rich interactions between the solid, liquid, and vapor phases \citep{quere2013leidenfrost}. Examples include the evaporation dynamics and geometry of the drop \citep{burton2012geometry,biance2003leidenfrost,myers2009mathematical,pomeau2012leidenfrost,xu2013hydrodynamics,sobac2014leidenfrost,hidalgo2016leidenfrost,maquet2016leidenfrost,wong2017non}, the stability of the vapor-liquid interface \citep{duchemin2005static,lister2008shape,snoeijer2009maximum,bouwhuis2013oscillating,trinh2014curvature,raux2015successive,maquet2015leidenfrost}, hydrodynamic drag-reduction \citep{vakarelski2011drag,vakarelski2012stabilization,vakarelski2014leidenfrost}, self-propulsion of droplets \citep{linke2006self,dupeux2011viscous,dupeux2011trapping,lagubeau2011leidenfrost,cousins2012ratchet,li2016directional,soto2016surfing,sobac2017self}, impact dynamics \citep{biance2006elasticity,tran2012drop,castanet2015drop,shirota2016dynamic}, green nanofabrication \citep{abdelaziz2013green}, chemical reactions \citep{bain2016accelerated}, fuel combustion \citep{kadota2007microexplosion}, quenching process in metallurgy \citep{bernardin1999leidenfrost}, heat transfer \citep{talari2018leidenfrost,shahriari2014heat}, directional transport \citep{li2016directional}, soft heat engines \citep{waitukaitis2017coupling} and thermal control of nuclear reactors \citep{van1992physics}.

In many of these examples, transient and self-sustained capillary oscillations play an important role in the dynamics. The interplay between gravity, the flow of vapor beneath the drop, and the liquid surface tension can lead to both small- and large-amplitude oscillations with very little damping. An understanding of these detailed interactions is crucial for the stability of the vapor layer, the failure of which can lead to explosive boiling upon contact with the hot surface. However, the excitation mechanism of these oscillations is complicated by the presence of both hydrodynamic and thermal effects, for example, rapidly-flowing vapor can cause a strong shear stress at the liquid-vapor interface, and temperature gradients in the liquid can lead to convective and Marangoni forces. Here we focus on capillary oscillations in individual Leidenfrost drops, where the shape is mostly determined by the competition between the gravity and surface tension, as measured by the relative size of the drop with respect to the capillary length, $l_c\equiv \sqrt{\gamma/\rho_l g}$, where $\gamma$ and $\rho_l$ denote the surface tension and density of the liquid, and $g$ is the acceleration due to gravity. For drops with radius $R<l_c$, surface tension forces are dominant, and the drop shape is essentially spherical except for a vanishingly small flat region near the solid surface \citep{burton2012geometry}. \citet{caswell2014dynamics} identified a planar vibrational mode (``breathing" mode) in the neck region of the drop closest to the solid substrate. The oscillation frequencies were found to obey a power law that is not consistent with a general three-dimensional dispersion relation for capillary waves \citep{rayleigh1879capillary}. In this paper, we provide an analytical expression for the breathing mode using a simple model based on gravity and surface tension which shows excellent agreement with the experimental data.

Large Leidenfrost drops form liquid puddles whose thickness is approximately 2$l_c$. These puddles are known to spontaneously form large-amplitude, star-shaped oscillations \citep{holter1952vibrations,adachi1984vibration,strier2000nitrogen,snezhko2008pulsating,strier2000nitrogen,ma2015many,Maleidenfrost2016}. Similar oscillations have been observed in large drops on periodically-shaken, hydrophobic surfaces \citep{noblin2005triplon,noblin2009vibrations}, drops levitated by an underlying airflow \citep{bouwhuis2013oscillating}, and drops excited by an external acoustic or electric field \citep{shen2010parametric,shen2010parametrically,mampallil2013electrowetting}. In studies where the frequency of external forcing is prescribed, the oscillations are excited by a parametric mechanism \citep{brunet2011star}. The external forcing leads to variations in the drop radius with time. Since the drop radius appears in the dispersion relation for azimuthal, star-shaped oscillations, the evolution of the oscillation amplitude obeys an equation similar to the Mathieu equation. For Leidenfrost drops, however, the mechanism is less clear since there is no prescribed frequency, and the star oscillations are excited and sustained through the heat input and resulting evaporation of the liquid. It has been suggested that the star oscillations may result from modulations of the surface tension of the liquid due to temperature variations \citep{adachi1984vibration,takaki1985vibration,tokugawa1994mechanism}, or perhaps due to convective patterns \citep{snezhko2008pulsating,strier2000nitrogen}. However, \citep{bouwhuis2013oscillating} observed star-shaped oscillations in drops which are supported by an external, steady air flow, suggest that a hydrodynamic coupling between the gas flow and liquid interface initiates the oscillations. 

Given the importance of the Leidenfrost effect in basic fluid and thermal transport, or the numerous practical applications, we know surprisingly little about the coupling between the evaporated vapor flow and vapor-liquid interface that lead to rich dynamical phenomena. Here we explore this coupling by investigating the self-organized, star-shaped oscillations of Leidenfrost drops using six different liquids: water, liquid N$_2$, ethanol, methanol, acetone and isopropanol. The liquid drops were levitated on curved surfaces in order to suppress the Rayleigh-Taylor instability, and star-shaped oscillation modes with $n$ = 2 to 13 lobes along the drop periphery were observed. The number of observed modes depended sensitively on the liquid viscosity, whereas the oscillation frequency (wavelength) depended only on the capillary length but not the mode number, substrate temperature, or drop size. Accompanying pressure measurements in the center of the vapor layer indicate that the pressure variation frequency was approximately twice that the drop oscillation frequency for all of the observed modes, consistent with a parametric forcing mechanism. We show that the pressure oscillations are driven by capillary waves of a characteristic wavelength beneath the drop traveling from the drop center to the edge, and such capillary waves can be generated by a strong shear stress at the liquid-vapor interface. Additionally, we find that although thermal convection is expected to be quite strong, the robust frequency (wavelength) of star oscillations is only weakly affected by varying either the substrate or environmental temperature, suggesting that star-shaped oscillations of Leidenfrost drops are hydrodynamic in origin.

\section{Experiment}
\label{experiment}
In the experiment, blocks of engineering 6061 aluminum alloy with dimensions of $\sim$ 7.6 cm $\times$ 7.6 cm $\times$ 2.5 cm were used as substrates. Resistive heaters were embedded with high-temperature cement into the aluminum to control the substrate temperature, $T_s$. Six different liquids were used as Leidenfrost drops: deionized water, liquid nitrogen (liquid N$_2$), ethanol, methanol, acetone, and isopropyl alcohol (isopropanol). The relevant liquid properties at the boiling point, $T_ {b}$, such as the surface tension $\gamma$, density $\rho_l$, dynamic viscosity $\eta_l$, and capillary length $l_c$, are listed in table \ref{tab:1}. The substrates were heated to different temperatures based on $T_ {b}$ for each liquid. For water, the temperature of the substrate was set from 523 K to 773 K, for ethanol, methanol, acetone, and isopropanol the temperature of the substrate was set to 523 K, whereas the substrate for liquid N$_2$ was not heated due to its extremely low $T_ {b}$. 

\begin{figure}
\begin{center}
\includegraphics[width=4 in]{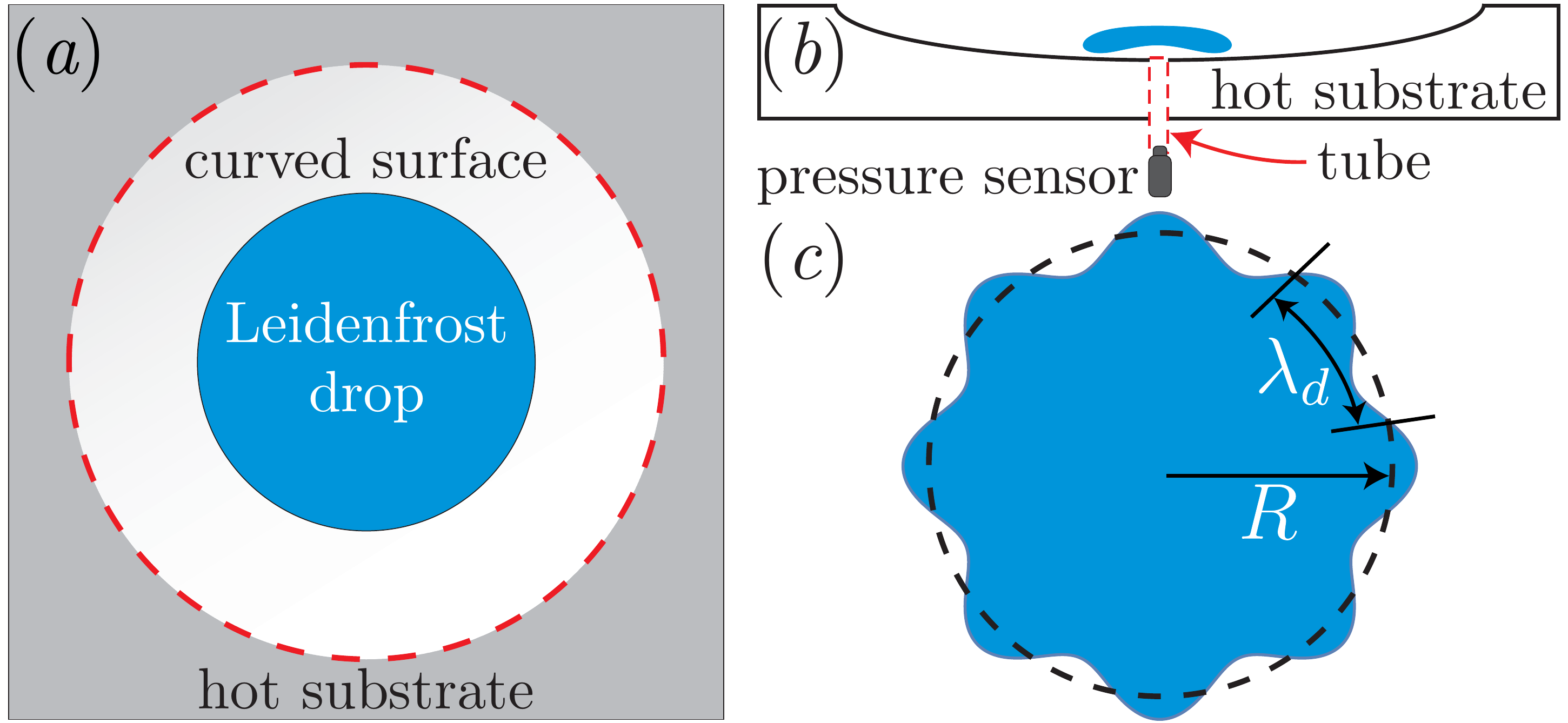}
\caption[]{(Color online) Schematics of the experimental setup and star-shaped oscillation pattern. (\textit{a}) Top-down view of the substrate whose upper surface is milled into a bowl shape enclosed by the dashed red circle. (\textit{b}) Cross-sectional view of the substrate. (\textit{c}) Star-shaped oscillation pattern of a Leidenfrost drop showing the standing wave along the drop periphery, which obeys $2\pi R=n\lambda_d$, where $R$ is the drop radius, $n$ is the number of lobes, and $\lambda_d$ is the wavelength of the standing wave.
} 
\label{setup}
\end{center} 
\end{figure}

The upper surfaces of the substrates were machined into a concave, spherical shape in order to suppress the buoyancy-driven Rayleigh-Taylor instability at the vapor-liquid interface and keep the drops stationary \citep{snoeijer2009maximum,quere2013leidenfrost,trinh2014curvature}. After machining, the roughness of the surface was inspected using an optical microscopy with 50$\times$ magnification. By changing the focus, we determined that over a 250 $\mu$m $\times$ 250 $\mu$m surface area, the peak-to-peak roughness was less than 10 $\mu$m, and often much smaller than this value. The curved surfaces for different liquids were designed to satisfy $l_{c}/R_s$ = 0.03, where $R_s$ is the radius of curvature of the surface whose top and cross-sectional views are schematically shown in figures \ref{setup}(\textit{a}) and \ref{setup}(\textit{b}), respectively. Following this principle, we fabricated three types of curved substrates, i.e. one for water, one for ethanol, methanol, acetone, and isopropanol, and one for liquid N$_2$, considering their respective capillary lengths ($l_c$) listed in table \ref{tab:1}. For some experiments, a plano-concave, fused silica lens (focal length = 250 mm) was used as the heated substrate in order to allow for optical imaging of the capillary waves underneath the Leidenfrost drop. A T-type thermocouple (maximum measurable temperature: 473 K, tip diameter $\approx$ 0.08 cm, HYP-2, Omega Engineering) was used to measure the internal temperature profile of Leidenfrost drops. 

For most of the substrates, a pressure sensor (GEMS Sensors, response time: 5 ms, sensitivity: 2 mV/Pa) was connected to a hole (diameter = 1 mm) at the center of the curved substrate in order to measure the pressure variations in the vapor layer at sample rates of 500-1000 Hz during quiescent and oscillatory phases of drops as illustrated in figure \ref{setup}(\textit{b}). We used a high-speed digital camera (Phantom V7.11, Vision Research) with a resolution of about 132 pixels/cm to image the motions of drops from above at frame rates of 1000 frames per second. Recorded videos were then analyzed with NIH ImageJ software to obtain the frequency and wavelength of the star-shaped oscillations (A typical star-shaped oscillation mode $n$ = 8 is schematically shown in figure \ref{setup}\textit{c}).

\begin{table}
\begin{center}
\def~{\hphantom{0}}
\begin{tabular}{lcccccccc}
     liquid                 &$T_{b}$  &$\gamma$      &$\rho_l$   &$\eta_l$   &$\l_{c}$  &$T_s$ &Modes &$Re_l$\\       
     \hline
     water                &373        &59.0	        &958	    &0.282	&2.5       &523-773 &2-13       &1340     \\
     liquid N$_2$	    &77          &8.90     	 &807	    &0.162	&1.1       &298 &3-5,7        &539      \\
     acetone	          &329       &18.2	       &727	    &0.242	&1.6       &523 &5-10       &601    \\
     methanol	          &338        &18.9	       &748	    &0.295	&1.6       &523 &6-10        &511      \\
     ethanol              &352	     &18.6	       &750	    &0.420	&1.6       &523 &7-11      &355        \\
     isopropanol        &356      &15.7	       &723	    &0.460 	&1.5      &523  &9,10	       &283   \\
     \hline 
\end{tabular}
\caption{Physical properties of different liquids at the boiling point $T_b$ (K). Units are as follows: $\gamma$ (mN/m), $\rho_l$ (kg/m$^3$), $\eta_l$ (mPa s), $l_c$ (mm). Data was taken from \citet{lemmon2011nist}. The last three columns indicate the range of substrate temperatures, $T_s$ (K), used in the experiments, the observed mode numbers ($n$), and the Reynolds number computed in \S\ref{stars of different liquids}.}
\label{tab:1}
\end{center}
\end{table}

\section{Results and discussions}
\label{results and discussions}
\subsection{The geometry of Leidenfrost drops on curved surfaces}
\label{dropshape}

The shape of a Leidenfrost drop is determined by a competition between surface tension and gravity. It has been shown that a nonwetting drop with radius $R \ll l_c$ will exhibit a quasi-spherical profile except for the bottom region where the drop is slightly flattened by gravity, which still holds true for small Leidenfrost drops \citep{burton2012geometry}. In comparison, for large Leidenfrost drops, i.e. $R \gg \l_c$, the shape of the drops is dominated by gravity and resembles a circular puddle with a constant thickness of approximately 2$l_c$ \citep{biance2003leidenfrost}. However, these large puddles are susceptible to the Rayleigh-Taylor instability, which is manifested by bubbles rising from the vapor layer beneath the drop \citep{biance2003leidenfrost}. In our experiment, we used curved surfaces (see \S\ref{experiment}) in order to suppress the Rayleigh-Taylor instability and obtain large, stable Leidenfrost drops \citep{snoeijer2009maximum,quere2013leidenfrost,trinh2014curvature}. 

\begin{figure}
\begin{center}
\includegraphics[width=3.5 in]{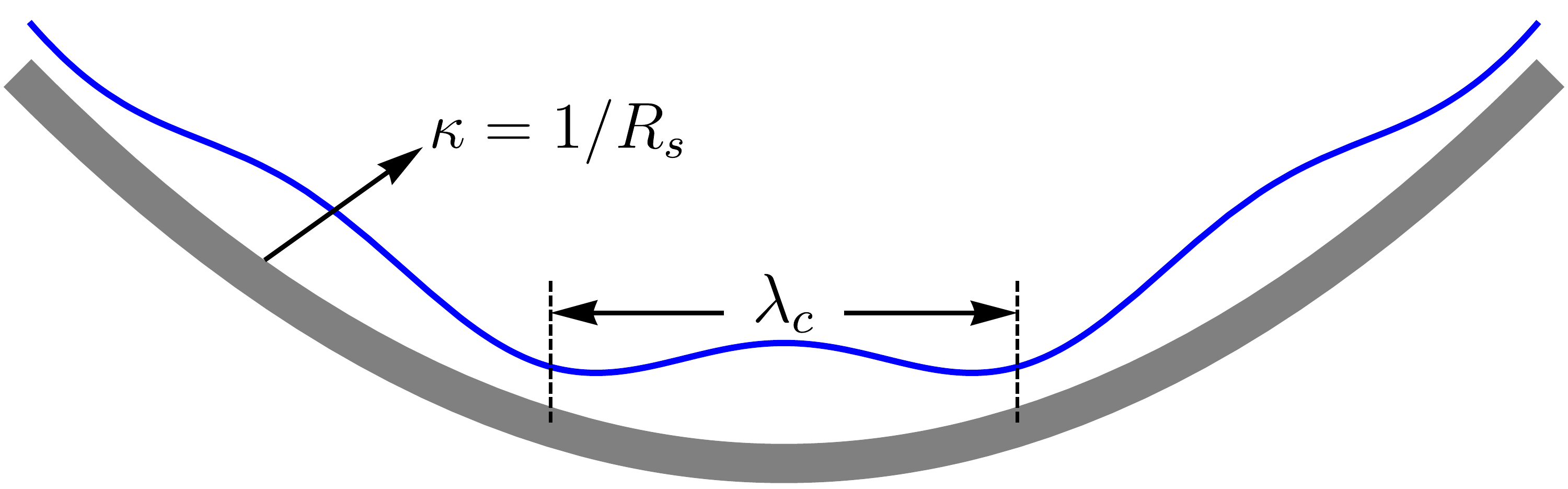}
\caption[]{(Color online) Schematics of the suppression of Rayleigh-Taylor instability at the liquid interface (solid blue curve) by a curved substrate (solid gray curve) with a curvature $\kappa=1/R_s$, where $R_s$ is the radius of curvature defined before, $\lambda_c$ is the capillary wavelength. 
} 
\label{RT_instability}
\end{center} 
\end{figure}

The effects of surface curvature on the Rayleigh-Taylor instability can be seen through the following simple model, similar to the original model used by \citet{biance2003leidenfrost}. Assuming the liquid-vapor interface beneath a Leidenfrost drop is perturbed with axisymmetric, sinusoidal variations, as illustrated in figure \ref{RT_instability}, then the new drop interface can be expressed in cylindrical coordinate as
\begin{equation}
S(r)=e+\frac{r^2}{2R_s}+\epsilon \cos \left( \frac{2\pi r}{\lambda_c} \right), 
\end{equation}
where $e$ is the mean vapor layer thickness, $\epsilon$ is the perturbation amplitude, and $\lambda_c$ is the capillary wavelength generated at the bottom of the drop (see figure \ref{sketch}). We also assume that $R_s\gg e\gg \epsilon$, implying a small perturbation to the equilibrium shape of the drop \citep{duchemin2005static,lister2008shape}. For a stable interface, the pressure at the drop center should be smaller or equal to the pressure at $r\approx\lambda_c/2$ in order to drive liquid back to the center. If the pressure in the vapor layer is constant, then to leading order, the pressures at $r$ = 0, and $r$ = $\lambda_c/2$ are
\begin{align}
\label{P0}
P_0\approx\gamma \kappa|_{r\rightarrow0}&=2\gamma \left( \frac{1}{R_s} -\frac{4\epsilon \pi^2}{\lambda_c^2}\right),\\
\label{Pneck}
P_1\approx\gamma \kappa|_{r\rightarrow \lambda_c/2}-\rho_l g\Delta z&=\frac{4\epsilon\pi^2 \gamma}{\lambda_c^2}+\frac{2\gamma+\rho_l g \lambda_c}{2R_s}-2\epsilon \rho_l g,
\end{align}  
where the height difference between the two points is $\Delta z=S(r)_{r\rightarrow0}-S(r)_{r\rightarrow \lambda_c/2}=2\epsilon-\lambda_c^2/8R_s$. The condition for stability can be found by equating (\ref{P0}) and (\ref{Pneck}), leading to an expression for the wavelength:
\begin{equation}
\lambda_c=2\left( 2\epsilon R_s-\frac{\sqrt{2\rho_l g\epsilon R_s(-3\pi^2\gamma +2\rho_l g\epsilon R_s)}}{\rho_l g} \right)^{1/2}.
\label{Rneck}
\end{equation}
Since the perturbation amplitude and radius of curvature always appear as a product, (\ref{Rneck}) can also be written as 
\begin{equation}
\dfrac{\lambda_c}{l_c}=2\left( 2\chi-\sqrt{2\chi(2\chi-3\pi^2 )} \right)^{1/2},
\label{Rneck_scale}
\end{equation}
where $\chi=\epsilon R_s/l_c^2$.  For finite $\epsilon$, in the limit $R_s\rightarrow\infty$, (\ref{Rneck}) simplifies to $\lambda_c/2\approx 3.85l_c$, in agreement with the prediction in \citet{biance2003leidenfrost} for a flat surface. However, the addition of a curved surface couples the perturbation amplitude to the radius of curvature. The quantity $2\chi-3\pi^2$ must be positive in order for $\lambda_c$ to be real, and thus represent the condition for instability. Subsequently, on a curved surface, the perturbation amplitude must satisfy $\epsilon\gtrsim14.8 l_c^2/R_s$ in order to lead to the Rayleigh-Taylor instability.

\begin{figure}
\begin{center}
\includegraphics[width=4 in]{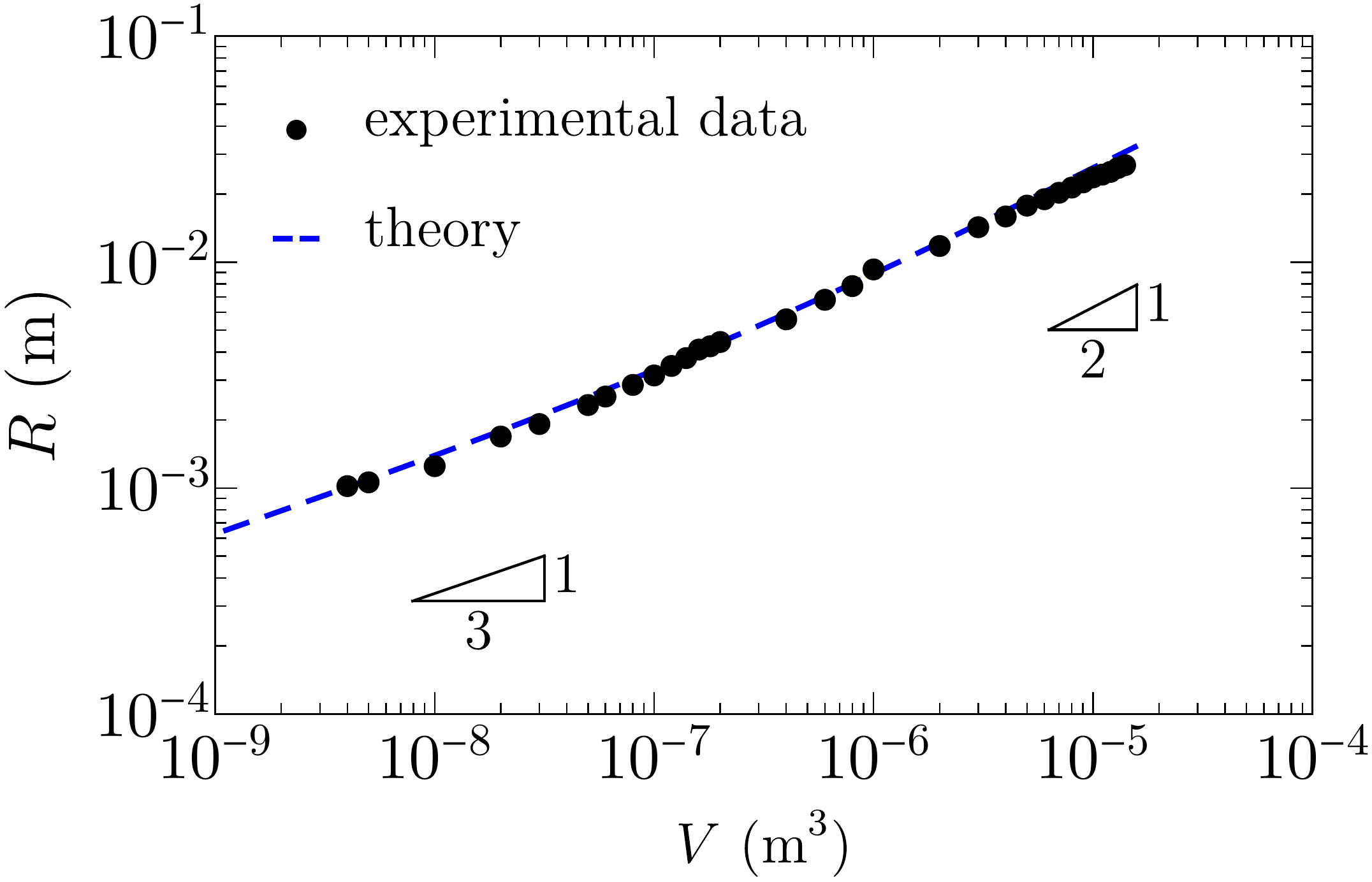}
\caption[]{(Color online) The variations of drop radius $R$ with respect to volume $V$ for Leidenfrost water drops. The dashed blue line shows the theoretical prediction for a drop on a flat surface assuming that the contact angle of the drop is $180^{\circ}$ as indicated in the text}. 
\label{RV}
\end{center} 
\end{figure}

Although the curved surface suppresses the Rayleigh-Taylor instability, it still has an affect on the overall drop shape. We investigated the influence of surface curvature on the drop shape by depositing water drops of given volumes $V$ on a curved surface with $T_s$ = 623 K. We measured the radius $R$ of the drops from recorded images taken immediately after deposition prior to the onset of oscillations. As can be seen in figure \ref{RV}, stable Leidenfrost drops with $R\approx 11l_c$ are obtained on the curved surface, while on a flat substrate, the maximum radius of a stable Leidenfrost drop is $R\approx3.95l_c$ \citep{snoeijer2009maximum,biance2003leidenfrost,burton2012geometry}, suggesting that surface curvature plays a crucial role in suppressing the Rayleigh-Taylor instability. The dashed blue line in figure \ref{RV} shows the theoretical prediction for a drop on a flat surface by solving the Young-Laplace differential equation numerically and assuming that the contact angle of the drop is $180^{\circ}$ \citep{burton2010experimental,burton2012geometry}. The experimental results show excellent agreement with the theoretical prediction, indicating that the shape of the Leidenfrost drops is not strongly affected by the surface curvature. More specifically, when $R<l_c$ (2.5 mm), the drops are quasi-spherical and thus $R\propto V^{1/3}$, whereas $R\propto V^{1/2}$ for $R>l_c$, which is expected for puddle-like drops with constant thickness. 

However, one can also notice that most of the experimental data is slightly less than the corresponding theoretical prediction, which we attribute to two possible reasons. First, for small drops, the evaporation begins from the moment of deposition on the hot surface, which removes a small amount of water prior to imaging. Second, large drops are thicker in the center due to the underlying curved substrate, leading to a smaller apparent radius for a given volume. Thus, although surface curvature and evaporation may somewhat reduce the drop size, the effects seem to be a minor influence on the overall drop shape.

\subsection{``Breathing mode" of small Leidenfrost drops}
\label{The breathing mode}

\citet{caswell2014dynamics} experimentally characterized axisymmetric oscillations in the radius of small Leidenfrost drops using interference imaging. The drops displayed small-amplitude changes in the radius of the flat region near the surface. Due to volume conservation, an increase (decrease) in the drop radius leads to a decrease (increase) in the thickness of the drop. \citet{caswell2014dynamics} found that the oscillation frequency of the breathing mode, $f_b$, obeyed a distinct power law, $f_b \propto R_0^{-0.68\pm0.01}$, where $R_0$ is the average drop radius during the oscillation. This dependence is distinctly different than the expected three-dimensional dispersion relation for inviscid spherical drops, $f\propto R_0^{-3/2}$ \citep{rayleigh1879capillary}. Here we provide an analytical model which explains this contrast and fits the experimental data with no adjustable parameters. 

To leading order, and due to the axisymmetry of the breathing mode, we model a Leidenfrost drop as an incompressible liquid cylinder of volume = $V$ and a time dependent radius, $R(t)$. We will assume that the bottom of the cylinder is fixed at $z=0$, which is reasonable if the thickness of the vapor layer varies much less than the radius. In cylindrical coordinates, $(r, \phi, z)$, the simplest form for the velocity which satisfies $\nabla\cdot\vec{\bf v}=0$ and the boundary conditions $\vec{\bf v}\cdot\hat{\bf r}|_{r=R(t)}=R'(t)$ and $\vec{\bf v}\cdot\hat{\bf z}|_{z=0}=0$ is:
\begin{equation}
\vec{\bf v}=\dfrac{R'(t)}{R(t)}\left(r,0,-2z\right).
\end{equation}
For simplicity, we will write $R(t)=R$, and $R'(t)=R'$. The total kinetic energy of the drop is then:
\begin{align}
T=&\int\frac{1}{2}\rho_l\left| \vec{\bf v} \right|^2\mathrm{d}V=\dfrac{\pi \rho_l R^2}{R'^2}\int_{0}^{R}\int_{0}^{h}r(r^2+4z^2)\mathrm{d}z\mathrm{d}r.
\end{align}
Evaluating the integrals, and using the fact that $V=\pi R^2h$, where $h$ is the time-dependent height of the cylinder, we obtain
\begin{align}
T=&\dfrac{1}{12}\rho_l V\left( 3+\frac{8V^2}{\pi ^2 R^6} \right)R'^2.
\end{align}
The total potential energy is the sum of gravitational potential energy and surface energy:
\begin{equation} 
U=\dfrac{1}{2}\rho_l Vgh+2\pi\gamma R(h+R)=\dfrac{gV^2\rho_l+4\pi\gamma R\left( V+\pi R^3 \right)}{2\pi R^2}.
\end{equation}  
\begin{figure}
\begin{center}
\includegraphics[width=3.5 in]{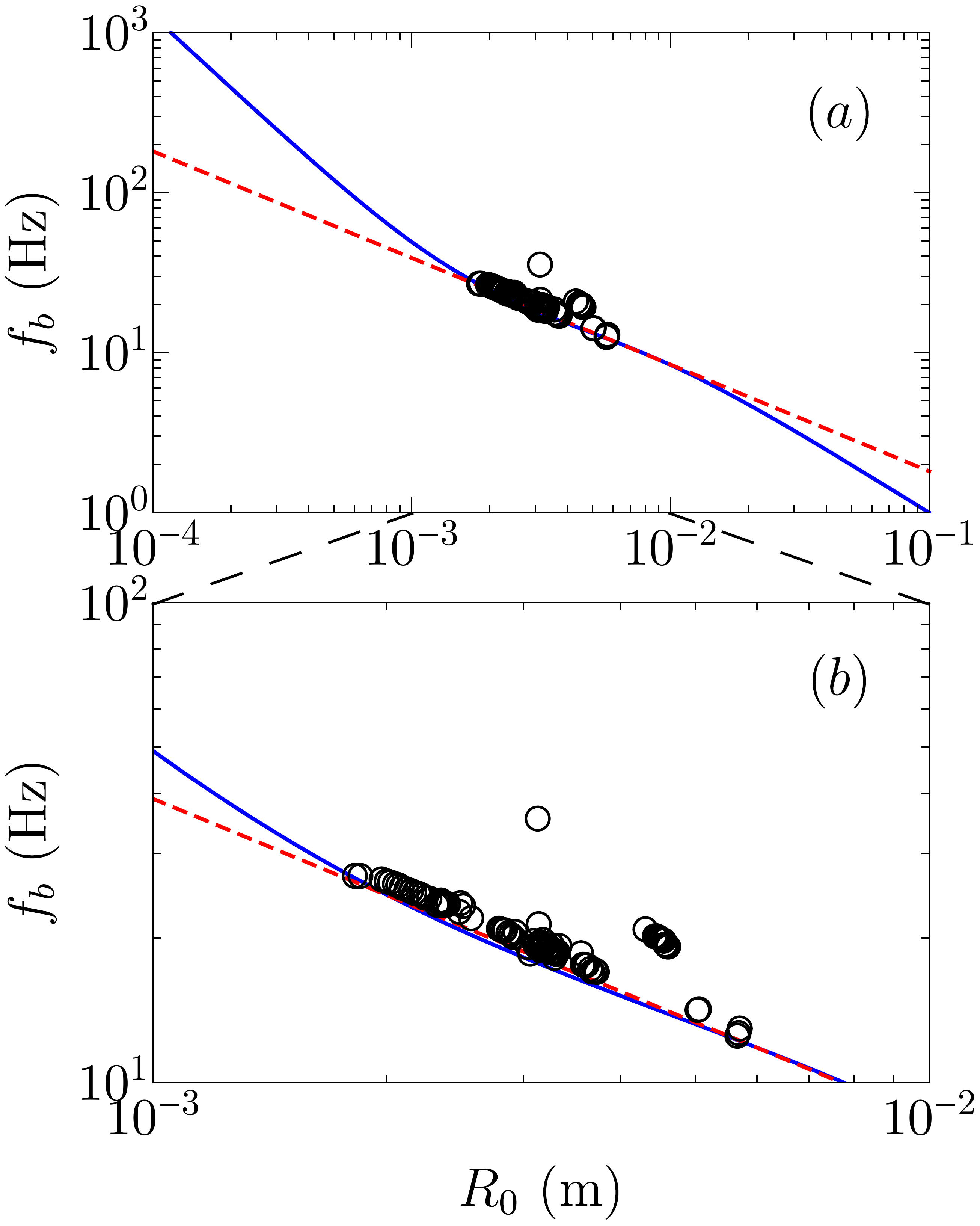}
\caption[]{(Color online)(\textit{a}) Breathing mode oscillation frequency $f_b$ of small Leidenfrost drops as a function of the average drop radius $R_0$. The dashed red curve represents the power law $f_b \propto R_0^{-0.68}$, and the solid blue curve represents the dispersion relation from (\ref{breathing_dispersion_relation}). (\textit{b}) A zoomed-in view of the data in (\textit{a}). With permission, the data shown here is taken from \citet{caswell2014dynamics}.
}        
\label{breathing_mode}
\end{center} 
\end{figure}
The equilibrium drop radius, $R_0$, can be found by minimizing the potential energy with respect to $R$,
\begin{equation} 
\dfrac{\mathrm{d}U}{\mathrm{d}R}=0=4\pi\gamma R_0-\dfrac{2 V \gamma}{R_0^2}-\dfrac{g \rho_l V^2}{\pi R_0^3}.
\label{eqR}
\end{equation}  
For large drops where $R\gg h$, the first and third terms must balance. Equating these two terms and using the fact that $V=\pi R^2 h$, we see that 
\begin{equation} 
h\approx2\sqrt{\dfrac{\gamma}{\rho_l g}},\hspace{12 pt}R_0\rightarrow\infty,
\label{heq}
\end{equation}  
which agrees with the expected asymptotic thickness of large Leidenfrost drops \citep{biance2003leidenfrost}. We may now define the Lagrangian of the system as $L=T-U$, and apply the Euler-Lagrange equation to obtain a differential equation for $R$:
\begin{equation} 
R''=\frac{6\rho_l V^2g\pi R^4+12\gamma V \pi^2R^5-24\gamma \pi^3 R^8+24\rho_l V^3R'^2}{8\rho_l V^3R+3\rho_l V\pi^2R^7}.
\label{R''} 
\end{equation}
To proceed further, we will linearize the equation by considering only small oscillations of the radius, $R=R_0(1+\epsilon e^{i \omega t})$, where $R_0$ is equilibrium drop radius found by solving (\ref{eqR}), $\epsilon$ is the perturbation amplitude, and $\omega$ is the angular frequency of the oscillation. Assuming that $\epsilon\ll 1$, to leading order (\ref{R''}) reduces to an expression for the angular frequency:
\begin{equation} 
\omega^2=\dfrac{576\gamma^2(\gamma +\Gamma)+36 \rho_l^2g^2  R_0^4 (\gamma +2 \Gamma )+384 \gamma \rho_l g  R_0^2 (5 \gamma +2 \Gamma )}{\rho_l  R_0^3 \left(1120 \gamma^2+9 \rho_l^2 g^2  R_0^4+192 \gamma  \rho_l g  R_0^2\right)},
\label{breathing_dispersion_relation}
\end{equation}
where $\Gamma\equiv\sqrt{\gamma^2+4\gamma \rho_l g R_0^2}$ and we have substituted $V$ for the equilibrium radius $R_0$ using (\ref{eqR}).

Taking typical values of the parameters in (\ref{breathing_dispersion_relation}) for water, we can plot the oscillation frequency of the breathing mode, $f_b=\omega/2\pi$, as a function of $R_0$, which is shown by the solid blue curve in figure \ref{breathing_mode}(\textit{a}). The theory shows excellent agreement with the data. For comparison, the power law $f_b \propto R_0^{-0.68}$ found by \citet{caswell2014dynamics} is plotted as the dashed red line in figure \ref{breathing_mode}(\textit{a}). A closer view of the comparison between the prediction by (\ref{breathing_dispersion_relation}) and $f_b \propto R_0^{-0.68}$ can be seen in figure \ref{breathing_mode}(\textit{b}). The data lie in the transition regime between two asymptotic limits, $R_0\ll l_c$ and $R_0\gg l_c$, which may explain the anomalous power law reported by \citet{caswell2014dynamics}. In these two limits, (\ref{breathing_dispersion_relation}) reduces to 
\begin{align} 
&\omega\approx\left(\dfrac{36\gamma}{35\rho_l R_0^3}\right)^{1/2},\hspace{12 pt}R_0\rightarrow 0\\
&\omega\approx\dfrac{4}{R_0}\left(\dfrac{g\gamma}{\rho_l}\right)^{1/4},\hspace{12 pt}R_0\rightarrow\infty\nonumber.
\end{align}  
The scaling for small drops is independent of $g$, which is expected based on their nearly-spherical shape. For large drops where $h\approx2l_c$ (\ref{heq}), the scaling is determined essentially by gravity: $\omega\sim\sqrt{2gh}/R_0$.

\begin{figure}
\begin{center}
\includegraphics[width=5 in]{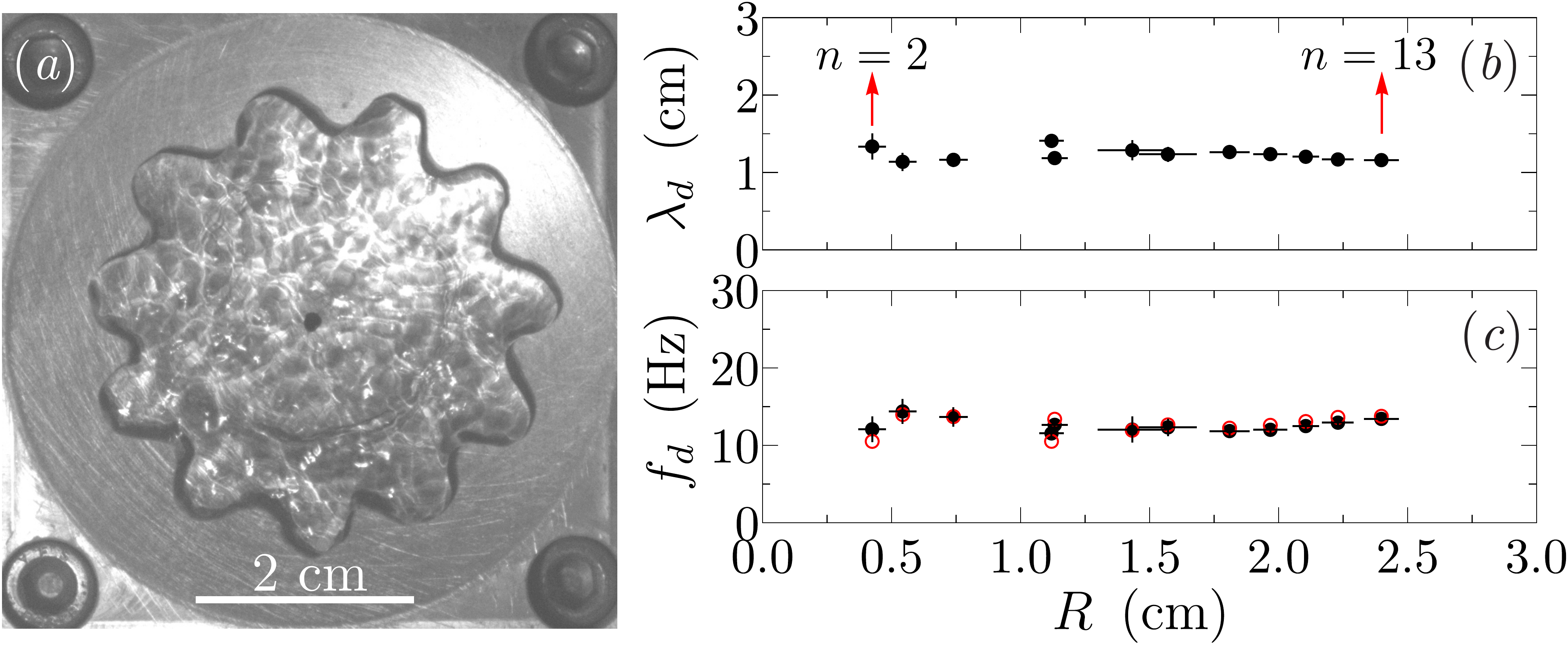}
\caption[]{(Color online) Star-shaped oscillations of Leidenfrost water drops at $T_s$ = 623 K. (\textit{a}) A snapshot of the oscillation mode with $n = 11$. (\textit{b}) and (\textit{c}) show the star-shaped oscillation wavelength $\lambda_d$, and frequency $f_d$, with respect to drop radius $R$. The data points from left to right  in (\textit{b}) and (\textit{c}) represent the increasing mode number $n$ as indicated by the red arrows, and the error bars in (\textit{b}) and (\textit{c}) are the standard deviations of multiple drops. The data for $\lambda_d$ in (\textit{b}) are indirectly measured using the relation $2 \pi R = n\lambda_d$, and the red circles in (\textit{c}) represent the theoretical prediction by (\ref{dispersion2D}) for drops with different radii.} 
\label{wavelength_frequency}
\end{center} 
\end{figure}

\subsection{Star-shaped oscillations of Leidenfrost water drops}
\label{stars of water drop}

In addition to the breathing mode, large Leidenfrost drops may also develop azimuthal, star-shaped oscillations. Similar oscillations have been observed in a variety of systems involving liquid drops \citep{brunet2011star}. In our experiments with Leidenfrost water drops using a curved substrate, we observe star-shaped oscillation modes with $n$ = 2 to 13 lobes along the drop periphery at $T_s=623$ K. A typical star-shaped oscillation mode ($n=11$) is shown in figure \ref{wavelength_frequency}(\textit{a}). Figures \ref{wavelength_frequency}(\textit{b}) and \ref{wavelength_frequency}(\textit{c}) show the drop oscillation wavelength $\lambda_d$ and frequency $f_d$ for different $R$, respectively, which are measured by analyzing the high-speed videos of the oscillating drops. Surprisingly, both $\lambda_d$ and $f_d$ remained nearly constant as we varied $R$. Increasing $R$ thus led to an increase in the allowable number of lobes $n$, as indicated by the red arrows. This similar trend also applies to other liquids used in our experiments, which will be discussed in \S\ref{stars of different liquids}.

For the free oscillations of an incompressible, axisymmetric spherical drop with infinitesimal deformations, the natural resonance frequency, $f_n$, of the $n$th-mode is given by \citep{rayleigh1879capillary}:
\begin{equation} 
f_n=\frac{1}{2\pi}\sqrt{\frac{n(n-1)(n+2)}{\rho_l R^{3}}}.
\label{dispersion3D} 
\end{equation}  
For a liquid puddle with $R\gg h\approx$ 2$l_c$, where $h$ is the thickness of the liquid puddle \citep{biance2003leidenfrost}, the resonance frequency $f_n$ takes the form \citep{yoshiyasu1996self}:
\begin{equation} 
f_n=\frac{1}{2\pi}\sqrt{\frac{n(n^{2}-1)}{\rho_l R^{3}}}\sqrt{{\frac{1}{1+(2-\frac{\pi}{2}+\frac{n-3}{4})\frac{l_c}{R}}}}.
\label{dispersion2D} 
\end{equation}  
The first term under the square root is for a strictly two-dimensional drop, whereas the correction factor, $1/\sqrt{1+(2-\frac{\pi}{2}+\frac{n-3}{4})\frac{l_c}{R}}$, is due to the quasi-two-dimensional nature of the puddle. 

As shown in figure \ref{wavelength_frequency}, for water drops where we observed $n$ = 2 modes, the average radius was $R\approx$ 4 mm, which is greater than $l_c$. Hence, it is reasonable to use  (\ref{dispersion2D}) to predict the oscillation frequency for all star-shaped modes in our experiments. The comparison between theory and experiment is shown in figure \ref{wavelength_frequency}(\textit{c}), which indicates an excellent agreement. For smaller values of $n$, both $\lambda_d$ and $f_d$ vary non-monotonically with $R$. There are a few potential reasons for this. One possibility is that nonlinear effects play a more significant role for smaller values of $n$ because the ratio of the mode amplitude to the drop radius is larger \citep{becker1991experimental,smith2010modulation}. However, the excellent agreement with the linear theory, i.e. (\ref{dispersion2D}), suggests that nonlinear effects may not be important. 

Instead, we suggest that the behavior may be related to the mode selection mechanism. For a given radius, either the frequency or the wavelength is preferentially selected by the excitation mechanism. Once $f_d$ or $\lambda_d$ is selected, (\ref{dispersion2D}) will determine the other. As will be shown in \S\ref{capillary_waves_sec}, there are strong capillary waves excited at the liquid-vapor interface beneath the drop. These waves produce pressure oscillations which parametrically couple to the star-shaped modes. A determination of the expected behavior of $f_d$ on $R$ is thus complicated by the physics of the vapor flow beneath the drop, nevertheless, we would expect the effects to be less variable for $R\gg\lambda_d$, which agrees with the data shown in figure \ref{wavelength_frequency}. 

\subsection{Pressure oscillations in the vapor layer}
\label{pressure oscillation}

It has been well-documented that star oscillations of large liquid drops can be initiated by a parametric forcing mechanism \citep{brunet2011star}. For instance, when a liquid puddle is deposited onto a superhydrophobic, vertically-vibrating substrate with a prescribed frequency, then the equation of motion of the drop can be described by an equation similar to Mathieu's equation, and star oscillations will be excited when the drop oscillation frequency is approximately half that the excitation frequency \citep{yoshiyasu1996self,brunet2011star}. In the experiment, we observed star oscillations of Leidenfrost water drops (e.g. see figure \ref{wavelength_frequency}\textit{a}) in the absence of an external excitation. Therefore, we hypothesize that the star oscillations are driven by fluctuations in the vapor layer pressure that supports the drop.

In order to explore the dynamics of pressure fluctuations in the vapor layer, we utilized a pressure sensor attached to the center of the curved substrate, schematically shown in figure \ref{setup}(\textit{b}). We measured the pressure variations in the vapor layer of water drops at $T_s$ = 623 K during the initiation of the star-shaped oscillations. We started recording the pressure immediately after the drops were placed on the substrate, and stopped after the well-defined star-shaped oscillations had been initiated for several seconds. The whole process lasted about 18.5 s. 

\begin{figure}
\begin{center}
\includegraphics[width=5.3 in]{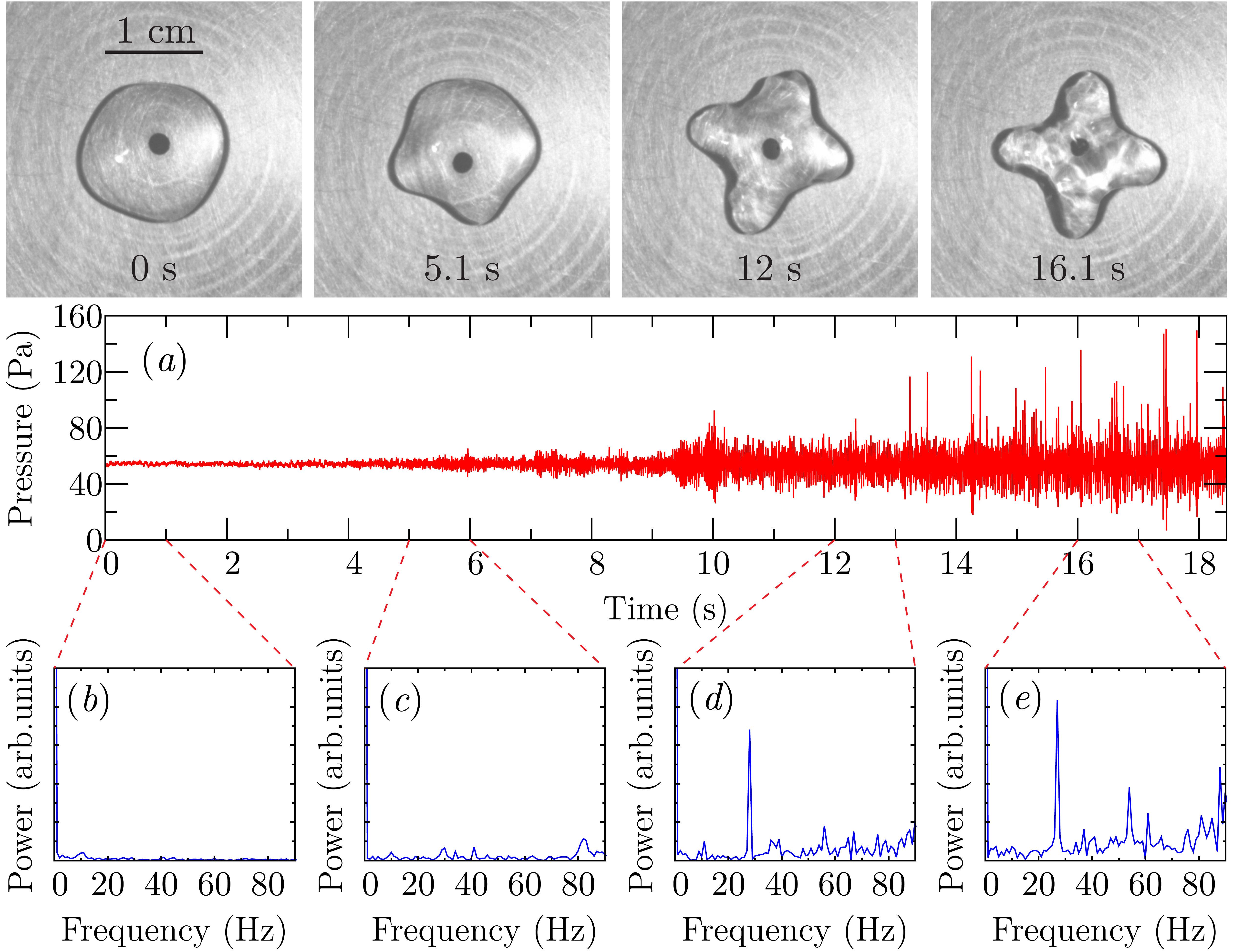}
\caption[]{(Color online) (\textit{a}) The pressure variations in the vapor layer during the initiation of a 4-mode star-shaped oscillation of a Leidenfrost water drop within $\approx$ 18.5 s.  (\textit{b}), (\textit{c}), (\textit{d}), and (\textit{e}) represent the Fourier power spectra of the pressure in the vapor layer during the time intervals of 0-1 s, 5-6 s, 12-13 s, and 16-17 s, respectively. The snapshots of the top panel represent the drop profile at 0 s, 5.1 s, 12 s, and 16.1 s, respectively, during the initiation process.} 
\label{parametric_oscillation}
\end{center} 
\end{figure}

The snapshots on the top panel of figure \ref{parametric_oscillation}(\textit{a}) show images of an $n$ = 4 drop at different stages during the oscillation. More specifically, the drop shape was nearly circular immediately upon placement on the substrate ($t$ = 0 s), then the pressure fluctuations in the vapor layer increased as the drop shape developed well-defined lobes. The amplitude continued to grow with time ($t$ =12 s) until the formation of a steady-state, star-shaped oscillation with large amplitude ($t$ = 16.1 s).  

Figure \ref{parametric_oscillation}(\textit{a}) shows the pressure variations in the vapor layer during the whole process. The mean pressure required to support the drop is $\rho_l gh\approx2\rho_l g l_c= 47$ Pa. However, the mean pressure we measure is slightly larger than this value because the pressure is measured at the center of the substrate where the the pressure is larger in order to drive the viscous vapor to flow out to the drop edge. The pressure fluctuations around the mean increase with time until the formation of a steady star-shaped oscillation with a large amplitude (about $t$ = 14 s).

In order to gain insight into the underlying relationship between the oscillations of the drop and pressure in the vapor layer, we performed a Fast Fourier Transform on the pressure data in four different time intervals: 0-1 s, 5-6 s, 12-13 s, and 16-17 s. The results are shown in figures \ref{parametric_oscillation}(\textit{b})-\ref{parametric_oscillation}(\textit{e}).  Initially, the pressure remains nearly constant, and there are no sharp peaks in the power spectrum (figure \ref{parametric_oscillation}\textit{b}). Between $t$ = 5-6 s, the pressure fluctuations become stronger and more periodic, and several small peaks are visible (figure \ref{parametric_oscillation}\textit{c}). Then, as the star oscillation is further developed, a sharp peak which is located at $\approx$ 28 Hz shows up in the power spectrum (figure \ref{parametric_oscillation}\textit{d}). Finally, when the star oscillation is fully developed, the pressure fluctuations stop growing, and a sharp peak located at $\approx$ 28 Hz dominates the spectrum (figure \ref{parametric_oscillation}\textit{e}). 

It is interesting to note that the location of the sharp peak is approximately twice the oscillation frequency of a fully-developed, $n$ = 4 mode Leidenfrost water drop (see figure \ref{wavelength_frequency}\textit{c}). This is consistent with a parametric forcing mechanism for the excitation of the star oscillations. Moreover, harmonics at higher frequencies are also visible as indicated by the secondary peaks in figure \ref{parametric_oscillation}(\textit{e}). This is likely due to nonlinear effects involved in the star-shaped oscillations. In addition, the location of the peak is robust. As will be shown in \S\ref{thermal effect}, the location of the dominant peak in the power spectrum is mostly independent of the substrate temperature and the environmental temperature. 

\begin{figure}
\begin{center}
\includegraphics[width=4 in]{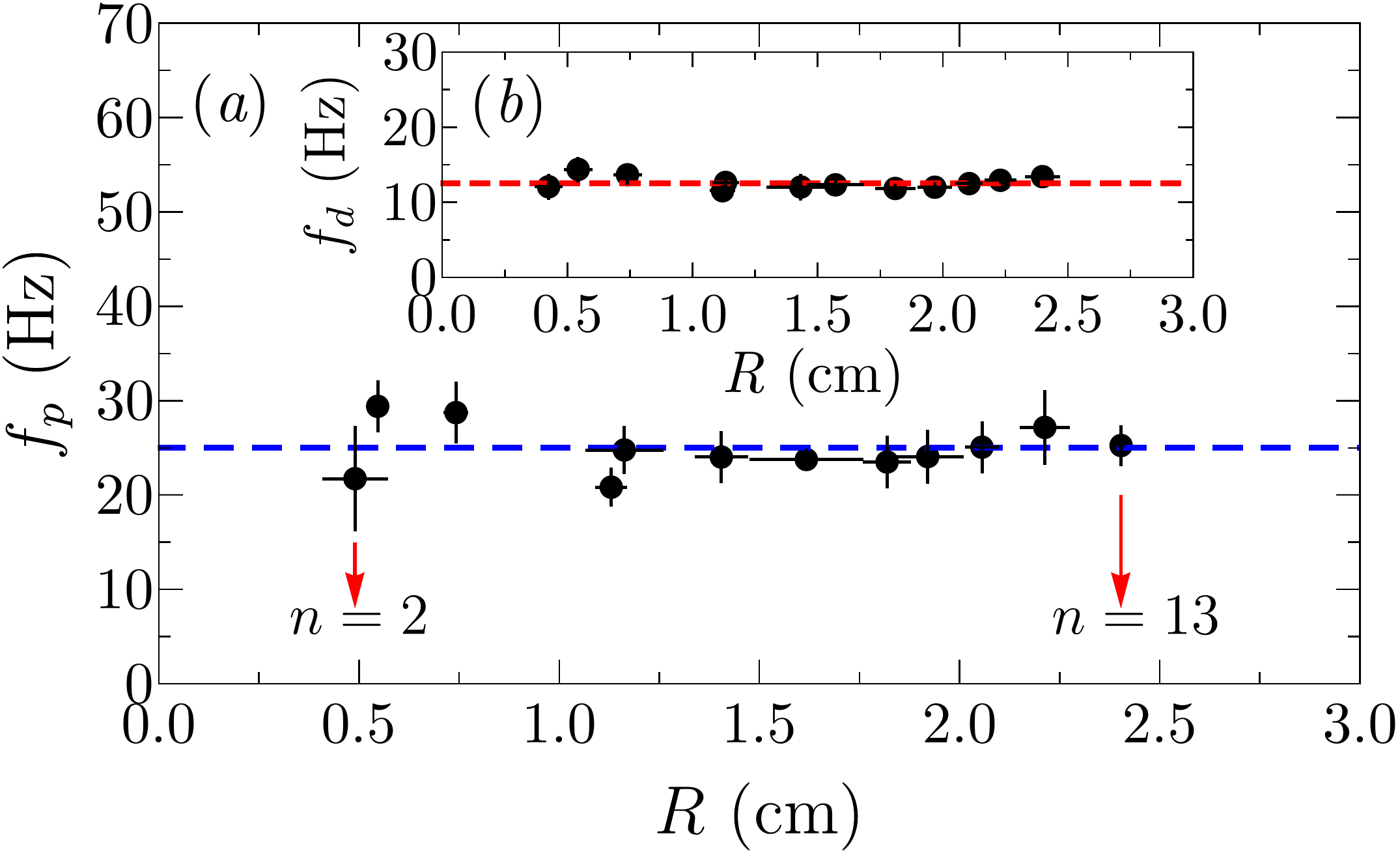}
\caption[]{(Color online) (\textit{a}) Summary of the pressure oscillation frequencies $f_p$ in the vapor layer of star-shaped Leidenfrost water drops with different radii at $T_s=623$ K. The data points from left to right represent the oscillation modes from $n=2$ to 13, respectively, as indicated by the red arrows. (\textit{b}) The drop oscillation frequency $f_d$ with respect to drop radius $R$ (figure \ref{wavelength_frequency}\textit{c}). The error bars of $f_p$ are taken from the full-width at half-max of the highest peak in the Fourier power spectra computed over a time interval of 10 s, and the error bars of $R$ are defined as the standard deviation of multiple measurements of different drops after the star oscillations are fully developed for each mode. The dashed lines in (\textit{a}) and (\textit{b}) are visual guides to indicate that $f_p\approx2f_d$.
} 
\label{f_vs_R}
\end{center} 
\end{figure}

Figure \ref{f_vs_R}(\textit{a}) shows the pressure oscillation frequency $f_p$ in the vapor layer of Leidenfrost water drops for all of the observed modes. The data points from left to right correspond the oscillation modes from $n$ = 2 to 13, as indicated by the red arrows. By comparing this with the drop oscillation frequency $f_d$ (figure \ref{f_vs_R}\textit{b}), we find $f_p\approx 2f_d$ as indicated by the dashed blue and red lines in figures \ref{f_vs_R}(\textit{a}) and \ref{f_vs_R}(\textit{b}), respectively. This robust relationship suggests that the star-shaped oscillations are parametrically driven by the pressure \citep{miles1990parametrically,kumar1994parametric,yoshiyasu1996self,brunet2011star,terwagne2011tibetan}. Consider the radial position of a point on the perimeter of the drop during a star-shaped oscillation, $r(t)$. This point will oscillate in time due to the azimuthal standing wave, and obey the following equation:
\begin{equation} 
\frac{\mathrm{d}^2r}{\mathrm{d}t^2}+\omega^2r=0.
\label{dropmotion}
\end{equation}  
To leading order, $\omega$ is the resonant frequency of the mode, (\ref{dispersion2D}):
\begin{equation}
\omega_{0}^{2}=\dfrac{n\left( n^2-1\right) \gamma}{\rho_l R^{3}}.
\label{dispersionrelationatR0}
\end{equation}
For simplicity, we have ignored the correction due to the quasi-two-dimensional nature of the drop. The pressure variations will induce vertical oscillations of the drop, and in turn oscillations of the drop radius:
\begin{equation} 
R=R_0(1+\epsilon \cos\omega_p t), 
\label{Rvariation}
\end{equation}  
where $R_0$ is the average radius of the drop, and $\epsilon$ is the amplitude of the small perturbation. Plugging (\ref{Rvariation}) into (\ref{dispersionrelationatR0}), to leading order in $\epsilon$ we obtain $\omega^2=\omega_0^2(1-3\epsilon\cos\omega_p t )$. Therefore, (\ref{dropmotion}) now becomes:
\begin{equation} 
\frac{\mathrm{d}^2r}{\mathrm{d}t^2}+\omega_0^2\left( 1-3\epsilon \cos \omega_p t \right)r=0.
\label{dropmotion1}
\end{equation}  

When $\omega_p \approx 2\omega_0$, the parametric resonance is the strongest, i.e. the oscillation amplitude will exponentially increase with time until the star-shaped oscillations are fully developed. This is consistent with our measurements of the pressure oscillation frequencies in the vapor layer (figure \ref{f_vs_R}), which are approximately twice that the drop oscillation frequencies (figure \ref{wavelength_frequency}\textit{c}) that we observed in our experiment. We note that it is possible that the pressure oscillations are a \emph{consequence} of the star oscillations. If the star oscillations were excited by some other mechanism, then the symmetry of the star shape would necessitate a pressure extrema when the star shape reaches its maximum amplitude. However, the only energy which is available to drive the oscillation comes from the evaporation and gas flow in the vapor layer. Thus we expect that the pressure oscillations are the source of the star oscillations. \S\ref{capillary_waves_sec} will explore the potential source of the pressure oscillations.

\begin{figure}
\begin{center}
\includegraphics[width=5.3 in]{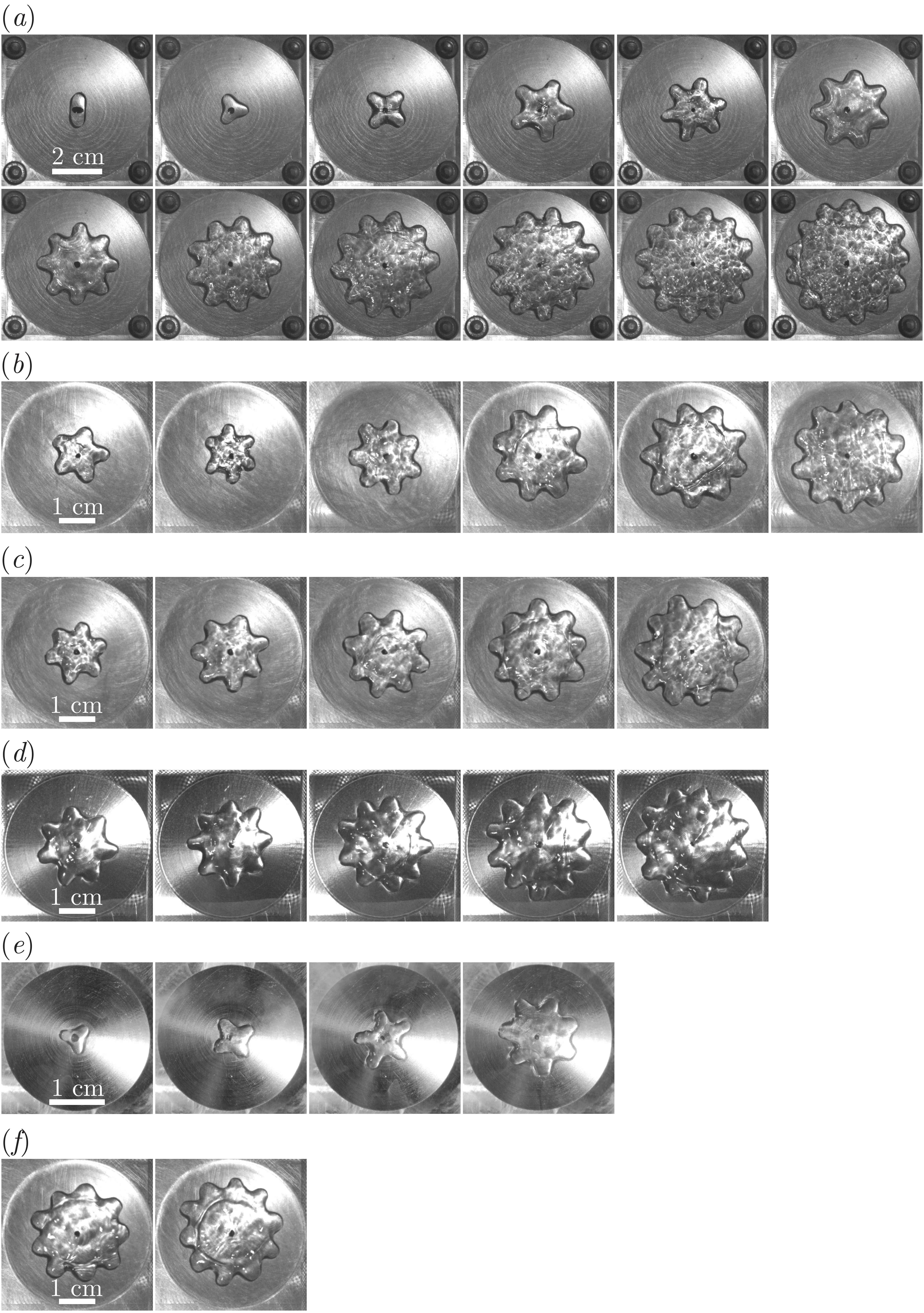}
\caption[]{(Color online) Snapshots of star-shaped oscillation modes with $n = 2$ to 13 of six different liquids when the lobes are at their maximum displacement. Panels (\textit{a}), (\textit{b}), (\textit{c}), (\textit{d}), (\textit{e}), and (\textit{f}) represent water, acetone, methanol, ethanol, liquid N$_2$, and isopropanol, respectively. For water: $T_s = 623$ K, for acetone, methanol, ethanol and isopropanol: $T_s = 523$ K. The substrate for liquid N$_2$ was not heated.
} 
\label{modegallery}
\end{center} 
\end{figure}

\subsection{Star-shaped oscillations of different liquids}
\label{stars of different liquids}

In addition to water, we performed similar experiments with five other liquids: liquid N$_2$, acetone, methanol, ethanol, and isopropanol. For these liquids, the curved surfaces were also machined to satisfy $l_{c}/R_s \approx 0.03$ (see \S\ref{experiment}), and the relevant physical properties and substrate temperatures are listed in table \ref{tab:1}. Figure \ref{modegallery} shows snapshots of star-shaped oscillations for the six different liquids we used in the experiments, where the other five liquids show similar star-shaped patterns to water drops but with different observable oscillation modes. By analyzing the images of the star-shaped oscillations, we find that star-shaped oscillations of acetone, methanol, ethanol, and isopropanol share a similar wavelength $\lambda_d \approx 0.9$ cm and frequency $f_d \approx 14$ Hz, whereas liquid N$_ 2$ shows a wavelength of $\lambda_d \approx 0.6$ cm and frequency $f_d \approx 17$ Hz. This suggests that the star oscillations may depend on the capillary length of the liquid.

\begin{figure}
\begin{center}
\includegraphics[width=3.5 in]{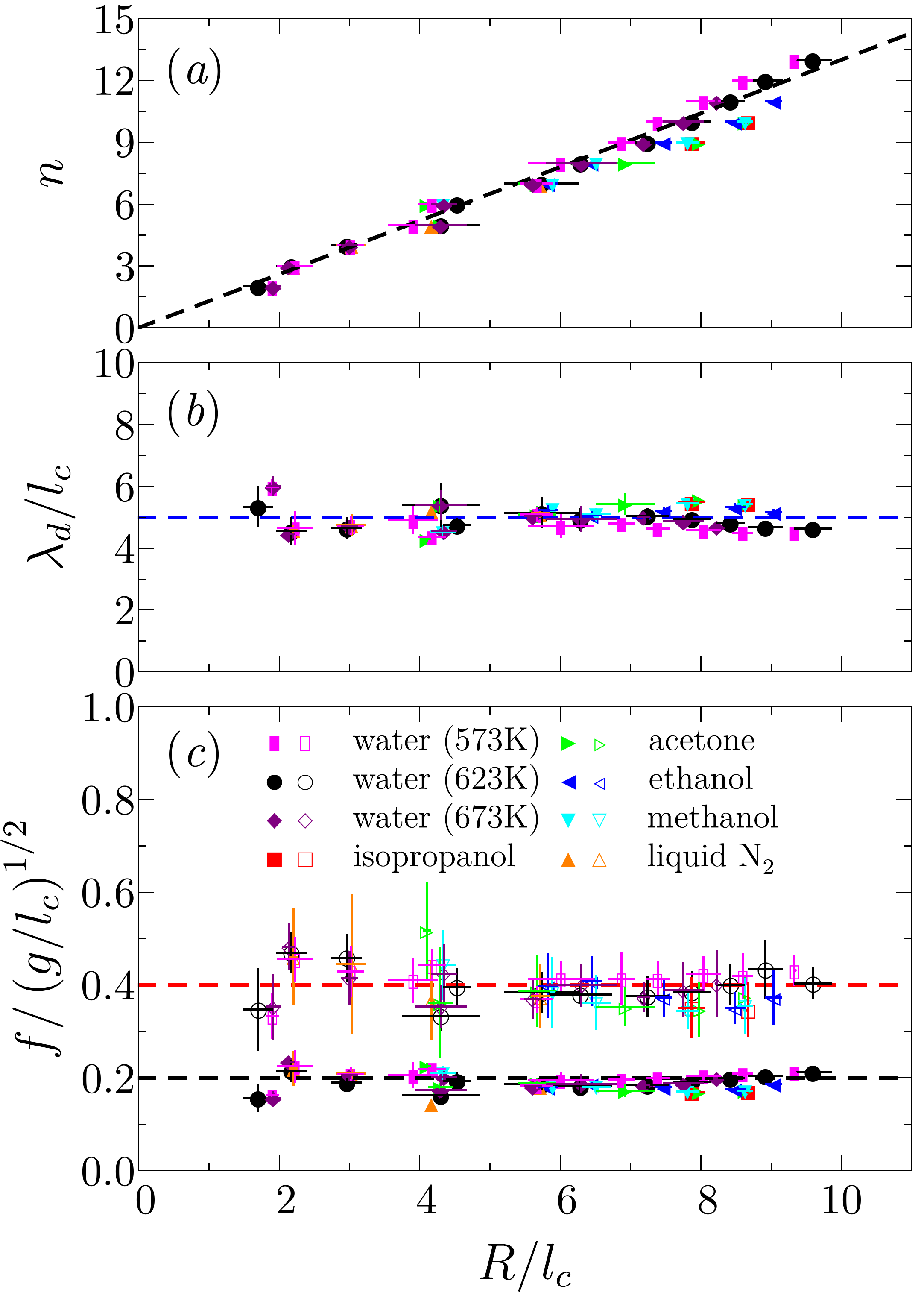}
\caption[]{(Color online) Dependence of star-shaped oscillation mode $n$ (\textit{a}), and dimensionless oscillation wavelength $\lambda_d/l_c$ (\textit{b}) and frequency $f/\sqrt{g/l_c}$ (\textit{c}) with respect to the dimensionless drop radius $R/l_c$ for six different liquid. In (\textit{c}), the solid symbols represent the drop oscillation frequency $f_d$, whereas the open symbols denote the corresponding pressure oscillation frequency $f_p$. The error bars of $R$, $\lambda_d$ and $f_d$ come from the standard deviations of multiple measurements of different drops, and the error bars of $f_p$ are defined as the full width of the highest peak in the power spectrum at half the maximum value. The dashed lines in all panels are visual guides.}
\label{scaling}
\end{center} 
\end{figure}

Figure \ref{scaling} shows the dependence of star-shaped oscillation mode $n$, rescaled wavelength $\lambda_d/l_c$, and rescaled frequency $f/\sqrt{g/l_c}$ on the rescaled drop radius $R/l_c$ of the six liquids used in the experiment. The mode number $n$ increases linearly with $R/l_c$ as indicated by the dashed black line in figure \ref{scaling}(\textit{a}), showing that large modes can only be observed in larger drops, as supported by figure \ref{modegallery}. The rescaled wavelength $\lambda_d/l_c$ for the six liquids collapses onto a straight line as shown in figure \ref{scaling}(\textit{b}), which indicates that the wavelength of the star oscillations only depends on $l_c$ of the liquid. Figure \ref{scaling}(\textit{c}) shows the oscillation frequency of both the azimuthal star oscillations, $f_d$, and the corresponding pressure, $f_p$, in the vapor layer. The solid symbols denote the drop oscillations, whereas open symbols with the same color represent the corresponding pressure oscillations in the vapor layer for a specific liquid. Both $f_d$ and $f_p$ collapse fairly well, and the relationship $f_p \approx 2f_d$ is consistent with the data, as indicated by the dashed black and red lines. Additionally, by comparing different values of $T_s$ for water, figure \ref{scaling} shows that $\lambda_d$, $f_d$, and $f_p$ are mostly independent of substrate temperature. 

The main influence of $T_s$ is the through the number of observable modes. For instance, at $T_s$ = 673 K, we did not observe oscillation modes $n$ = 12 and $n=13$ in water, which we attribute to the extremely fast evaporation rate of the liquid, as will be discussed in \S\ref{capillary_waves_sec}. Although the robust relationship between $f_d$ and $f_p$ suggests that the star oscillations of the drops are parametrically driven by the pressure in the vapor layer, there is an interesting variation in both $f_d$ and $f_p$ at small mode numbers. The data for smaller modes (e.g. $n= 2, 3, 4$) deviates further from the average compared to larger modes. This behavior may be nonlinear in nature since the amplitude of the mode relative to the drop radius is much larger for small $n$ \citep{becker1991experimental,smith2010modulation}. In addition, the quasi-two-dimensional nature of the drop is more important at small $n$, which may also contribute to the variations in this regime. A more quantitative explanation for this dependence is left for future studies.

\begin{figure}
\begin{center}
\includegraphics[width=4 in]{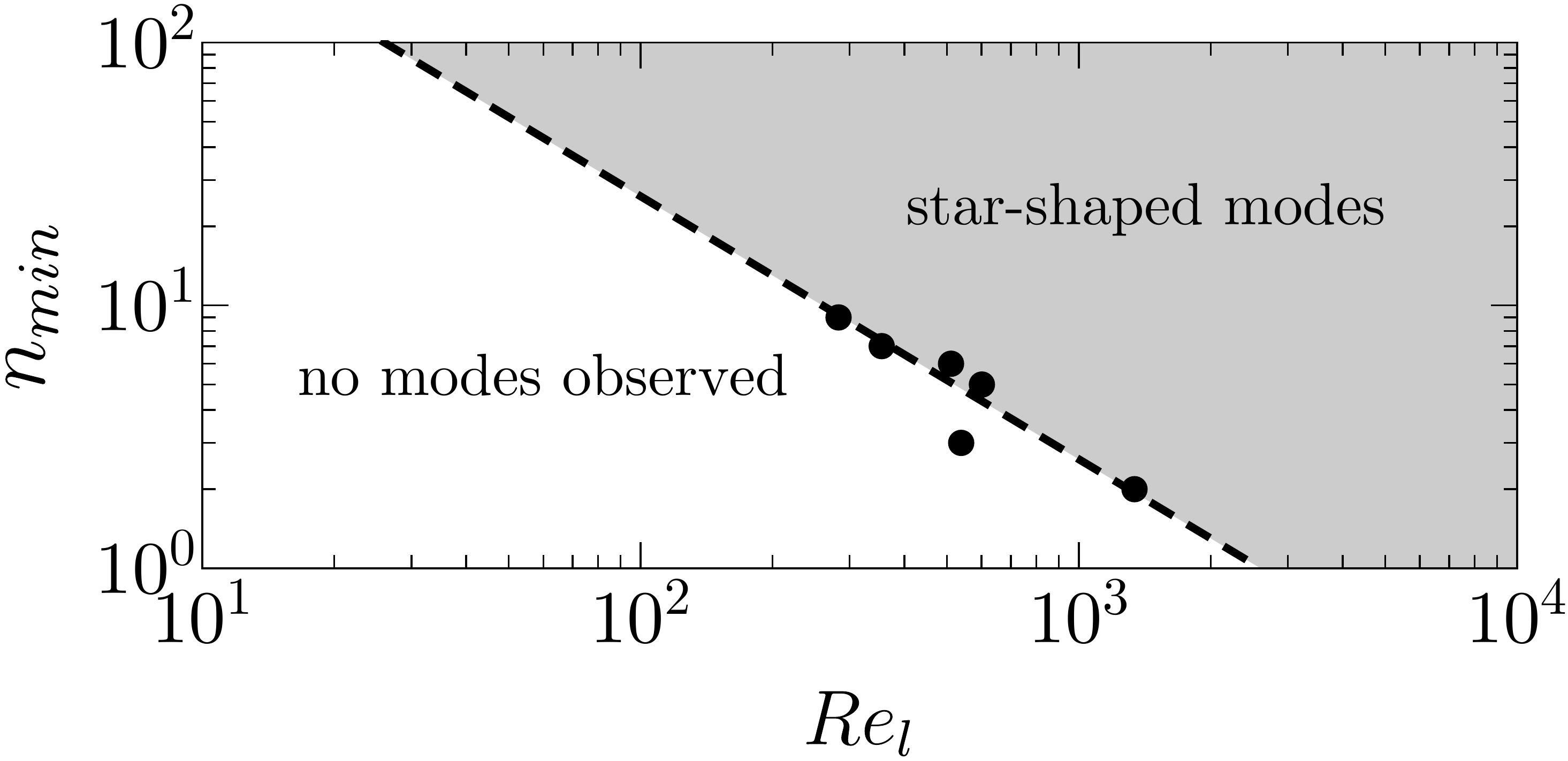}
\caption[]{Scaling behavior for the minimum observed star mode $n_{min}$ with respect to $Re_l$ of different liquids (table \ref{tab:1}). The dashed line represents the fit to the data, 2600 $Re^{-1}_l$.
} 
\label{threshold}
\end{center} 
\end{figure}

As shown in table \ref{tab:1}, different oscillation modes are observed in different liquids. In particular, some liquids failed to display smaller mode numbers. This may be due to inherent viscous damping that prevents the sustained excitation of large-amplitude oscillations. Following the analysis in \citet{Maleidenfrost2016}, the role of damping can be characterized by the Reynolds number associated with the liquid oscillation. For a star oscillation, the characteristic length scale and time scale are $l_c$ and $\sqrt{l_c^3\rho_l/\gamma}$, respectively. Thus, the inertial term in the Navier-Stokes equation is estimated as $|\rho_l(\vec{\bf v}\cdot\nabla)\vec{\bf v}|=\gamma/l_c^2$, and the viscous term is $|\eta_l\nabla^2\vec{\bf v}|=\eta_l/\sqrt{l_c^5\rho_l/\gamma}$. The Reynolds number of the oscillating liquid is defined as the ratio of inertial to viscous terms: $Re_l=\sqrt{l_c\rho_l\gamma}/\eta_l$, and the values of $Re_l$ for different liquids are listed in table \ref{tab:1}. 

Figure \ref{threshold} shows the dependence of the minimum mode number $n_{min}$ on $Re_l$ for each liquid. The dashed line represents a suggested scaling $Re^{-1}$, indicating that the liquid viscosity damps the oscillations of smaller drops and thus sets the minimum mode number of stable star oscillations. However, the substrate temperature and thus evaporation rate will likely affect the number of observed modes as well. A higher evaporation rate will induce a stronger driving of the oscillation modes, which would reduce $n_{min}$. This can be seen in figure \ref{threshold} for liquid N$_{2}$, which undergoes much more rapid evaporation due to the large difference between the boiling and substrate temperature, $T_s-T_b$. In addition, as mentioned previously, the stronger evaporation may also inhibit larger modes if the pressure oscillations are less coherent. In the next section we show how the flow in the vapor layer is linked to the pressure oscillations.

\subsection{Origin of pressure oscillations in the vapor layer}
\label{capillary_waves_sec}

As evidenced by figure \ref{scaling}(\textit{c}), the star-shaped oscillations of Leidenfrost drops are parametrically driven by the pressure variations in the vapor layer. Star oscillations induced by a parametric coupling have been observed in a variety of systems where variations of the drop radius are induced by external, periodically-modulated fields \citep{brunet2011star}. However, in our experiment, there are no obvious external fields. It is then crucial to understand the source of the pressure oscillations. In this section we show how capillary waves of a characteristic wavelength at the liquid-vapor interface induce the pressure oscillations in the vapor layer.

\begin{figure}
\begin{center}
\includegraphics[width=4 in]{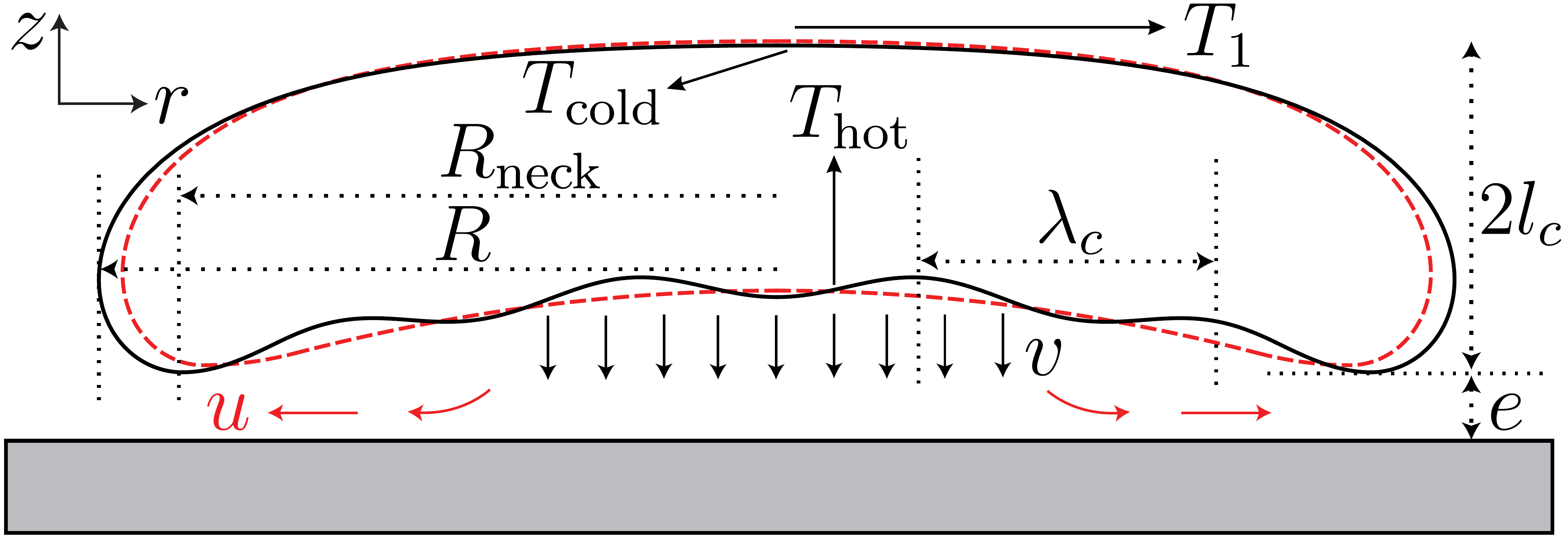}
\caption[]{(Color online) Sketch of the cross-sectional view of a large Leidenfrost drop levitating on a flat hot surface. The red dashed line and black solid line represent the drop profiles before and after the initiation of capillary waves at the bottom surface of the drop. The symbols are defined in the main text.
} 
\label{sketch}
\end{center} 
\end{figure}

Figure \ref{sketch} shows s sketch of a large, axisymmetric Leidenfrost drop with a maximum radius = $R$ and thickness $\approx2l_c$. The mean vertical velocity of the gas at the liquid surface is $v$, $e$ is the mean thickness of the vapor layer, and $u$ is the outward radial velocity of the gas near $r = R$. For such a large Leidenfrost drop, both the bottom surface of the drop and the substrate surface are assumed to be approximately flat to simplify the analysis. Following the model of \citet{biance2003leidenfrost}, the mass loss rate of the drop due to evaporation can be expressed as:
\begin{equation} 
\frac{\mathrm{d}m}{\mathrm{d}t}=\frac{\kappa_v}{L}\frac{\Delta T}{e}\pi R^2=\rho_v\pi R^2 v,
\label{evaeq1}
\end{equation}  
where $\kappa_v$ is the thermal conductivity of the vapor, $L$ is the latent heat of the evaporation, $\Delta T=T_s-T_b$, and $\rho_v$ is the density of the vapor at the liquid interface.  Since $e\approx$ 100 $\mu$m, and thus $R\gg e$ \citep{burton2012geometry}, we employ lubrication theory, leading to an expression relating the flow rate to the mean pressure, $P$, in the vapor layer:
\begin{equation} 
\frac{\mathrm{d}m}{\mathrm{d}t}=\rho_v\frac{2\pi e^3}{3\eta_v}P=\rho_v\frac{2\pi e^3}{3\eta_v}2l_c\rho_l g=\rho_v\pi R^2v,
\label{evaeq2}
\end{equation}  
where $\eta_v$ is the dynamic viscosity of the vapor. We note here that parameters such as $\kappa_v$, $\rho_v$, and $\eta_v$ will vary in the vapor layer due to the temperature gradient between the liquid and solid surfaces. For simplicity, we will assume that these represent average values of the vapor layer properties. A more detailed treatment, possibly including full numerical simulations, would be necessary to extend this simplified model.

In addition, due to mass conservation, the volumetric flow rate of gas from the bottom of the drop must be equal to the flow rate exiting the perimeter of drop:
\begin{equation} 
\pi R^2v=2\pi eRu.
\label{massconservation}
\end{equation}  

Solving (\ref{evaeq1}), (\ref{evaeq2}), and (\ref{massconservation}) for $u$, $v$, and $e$ yields:
\begin{eqnarray}
\label{u}
u=\left( \frac{\rho_l g\kappa_v l_c}{3\rho_v\eta_vL} \right)^{1/2}\Delta T^{1/2},\\
\label{v}
v=\left[ \frac{4l_c\rho_l g}{3\eta_v}\left( \frac{\kappa_v \Delta T}{L \rho_v} \right)^{3} \right]^{1/4}R^{-1/2},\\
\label{e}
e=\left( \frac{3\kappa_v \Delta T \eta_v}{4Ll_c\rho_l \rho_vg} \right)^{1/4}R^{1/2}.
\end{eqnarray}  

In this model, the steady-state, linear temperature profile in the vapor layer is valid since the characteristic time scale associated with thermal diffusion, $e^2/D_v\approx$ 1 ms, is smaller than the typical residence time of the gas in the vapor layer, $R/u\approx$ 10 ms, where $D_v$ is the thermal diffusivity of the vapor. Let us assume a water Leidenfrost drop with $R$ = 0.01 m, and for the vapor, we use the properties of steam at 373 K and 1 atm of pressure: $\rho_l$ = 959 kg/m$^3$, $\rho_v$ = 0.45 kg/m$^3$, $\gamma$ = 0.059 N/m, $g$ = 9.8 m/s$^2$, $L$ = 2.26 $\times 10^6$ J/kg, $\eta_v$ = 1.82 $\times$ 10$^{-5}$ Pa s, and $\kappa_v$ = 0.04 W/m/K. Although the properties of the vapor may vary somewhat in the vapor layer due to the temperature gradient, we have verified that this does not significantly affect our analysis. 

\begin{figure}
\begin{center}
\includegraphics[width=5 in]{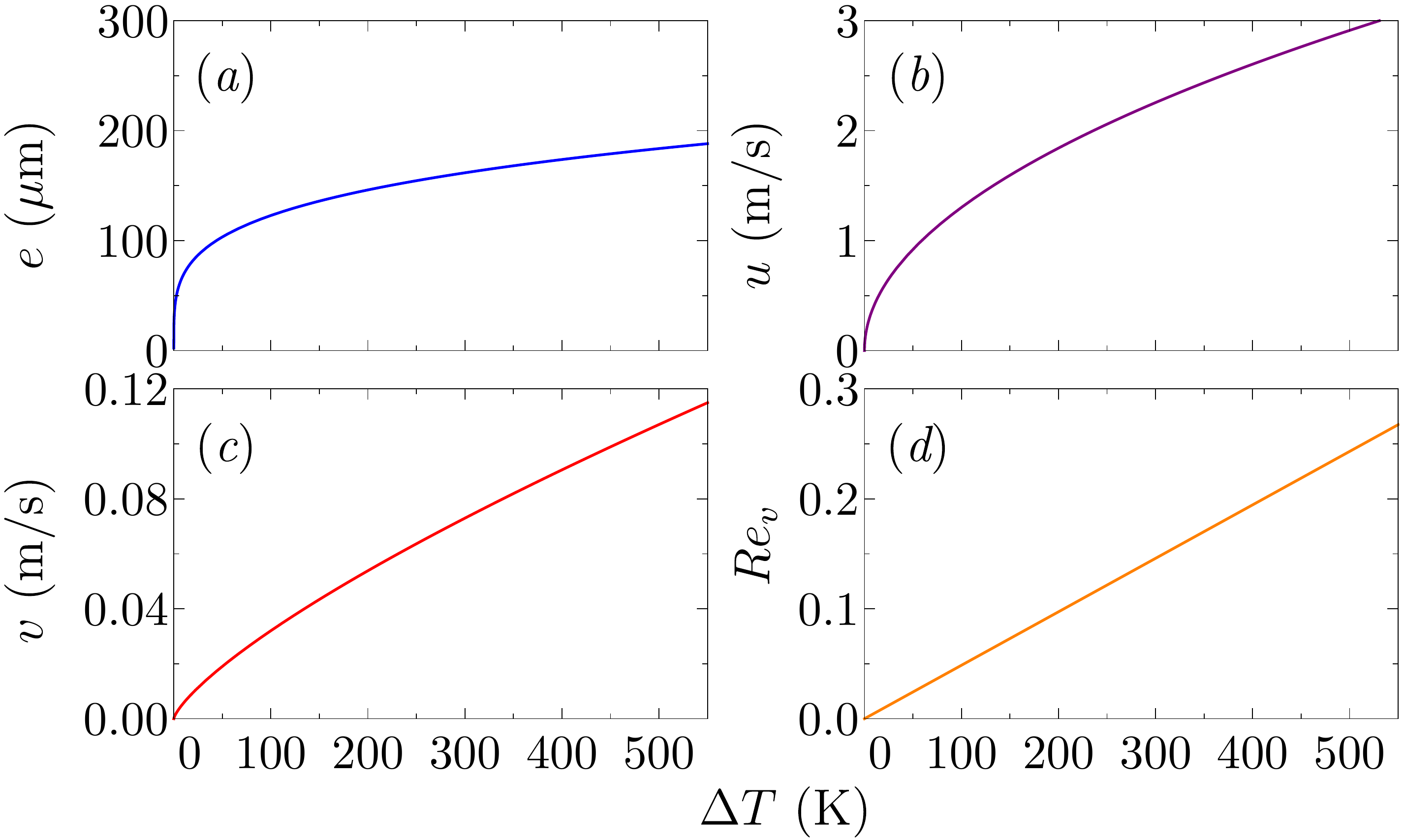}
\caption[]{(Color online) Dependence of vapor film thickness (\textit{a}), radial velocity (\textit{b}), vertical velocity (\textit{c}), and the Reynolds number in the vapor (\textit{d}) on temperature difference $\Delta T$, respectively.
} 
\label{vaporflow}
\end{center} 
\end{figure}

We can now plot $u$, $v$, and $e$ with respect to $\Delta T$ based on (\ref{u}), (\ref{v}) and (\ref{e}). The results are shown in figures \ref{vaporflow}(\textit{a}), \ref{vaporflow}(\textit{b}), and \ref{vaporflow}(\textit{c}), respectively. The inertial term in the Navier-Stokes equation is estimated as $|\rho_v(\vec{\bf v}\cdot\nabla)\vec{\bf v}|=\rho_vv^2/e$, and the viscous term as $|\eta_v\nabla^2\vec{\bf v}|=\eta_vv/e^2$. Thus the Reynolds number in the vapor layer is $Re_v=\rho_vve/\eta_v$. Plugging in (\ref{v}) and (\ref{e}), we arrive at a surprisingly simple expression for the Reynolds number in the vapor: $Re_v=\Delta T\kappa_v/L\eta_v$, which is independent of the drop size. This expression is plotted in figure \ref{vaporflow}(\textit{d}). Under typical experimental conditions, $Re_v\approx 0.2$, suggesting that the original lubrication flow assumption is valid, although inertial forces are non-negligible.

\begin{figure}
\begin{center}
\includegraphics[width=4.5 in]{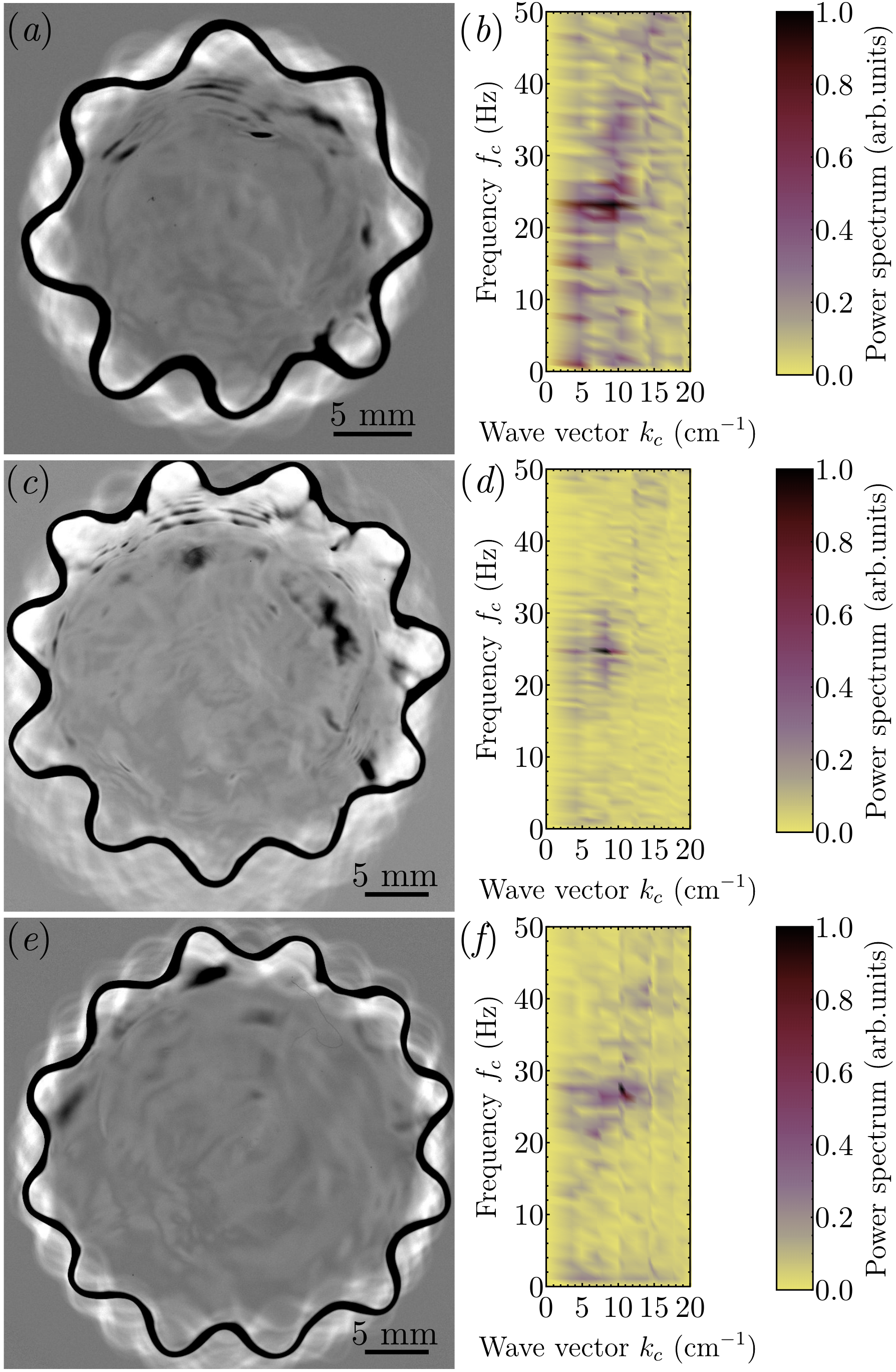}
\caption[]{(Color online) (\textit{a}) (\textit{c}) and (\textit{e}) show the capillary waves imaged beneath 8-mode acetone drop, 11-mode ethanol drop and 12-mode ethanol drop, respectively. (\textit{b}), (\textit{d}) and (\textit{f}) represent the corresponding Fourier power spectra of the capillary waves shown in (\textit{a}), (\textit{c}), and (\textit{e}), respectively. The images of capillary waves were enhanced for visibility, as described in the text. 
\label{capillary_waves}
} 
\end{center} 
\end{figure}

Based on figure \ref{vaporflow}, we can estimate the Bernoulli pressure in the vapor layer of Leidenfrost water drops as $\rho_vu^2/2\approx 1$ Pa, which is much smaller than the pressure variations $\approx$ 10 Pa (see figure \ref{parametric_oscillation}\textit{a}), suggesting that the inertial force does not account for the pressure variations. The viscous pressure is $P_{\mathrm{vis}}\sim \eta_vvR^2/e^3$, thus the pressure variation due to the local variations of the film thickness is $\Delta P_{\mathrm{vis}}=\frac{\mathrm{d}P_{\mathrm{vis}}}{\mathrm{d}e} \Delta e$. Then $\Delta P_{\mathrm{vis}}=$ 10 Pa corresponds to $\Delta e \approx$ 15 $\mu$m considering the typical values of $v$ (figure \ref{vaporflow}\textit{c}) and $R$. Although this local film thickness could also be possibly induced by the vertical motion of the center of mass of the drops, the fact that small and large drops (see figure \ref{scaling}\textit{c}) share a nearly constant oscillation frequency suggests this is not the case. Therefore, the pressure variations in the vapor layer are likely to be induced by the local variations of the vapor film thickness.

We propose that capillary waves with a characteristic wavelength, $\lambda_c$, traveling from the center to the edge of the drop lead to local variations in vapor film thickness, and thus pressure variations in the vapor layer, as schematically shown in figure \ref{sketch}. To confirm this possibility, we used a heated, plano-concave, fused silica lens as a substrate to image the capillary waves from below. The results for acetone and ethanol Leidenfrost drops are shown in figure \ref{capillary_waves}. The images were produced by averaging all frames in a given video sequence, then subtracting this background image in order to enhance contrast in the center of the drop. The white ``halo" surrounding the drop is a consequence of this subtraction process. The Fourier spectra for the capillary waves in the central region of the drops were computed in both time and space using the pixel intensity as the signal. Using the Nyquist sampling theorem, the maximum frequency was limited by half the video frame rate (500-1000 Hz), and the maximum spatial frequency was set by half the camera magnification (132 px/cm).

We observed a sharp peak in the measured Fourier spectrum for all the analyzed video sequences. In figures \ref{capillary_waves}(\textit{a}) and \ref{capillary_waves}(\textit{b}) (8-mode, acetone), this peak is located at a capillary wave frequency $f_c\approx$ 24 Hz and within a range of wave numbers, $k_c=2\pi/\lambda_c$, from 7 cm$^{-1}$ to 12 cm$^{-1}$, where $\lambda_c$ is the wavelength of the capillary waves. Figures \ref{capillary_waves}(\textit{c}) and \ref{capillary_waves}(\textit{d}) show the capillary waves beneath a 11-mode Leidenfrost ethanol drop and a sharp peak at a capillary wave frequency $f_c\approx$ 25 Hz with $k_c\approx$ 7-10 cm$^{-1}$. Similarly, for the capillary waves beneath a 12-mode Leidenfrost ethanol drop (figure \ref{capillary_waves}\textit{e}), a sharp peak exists at $f_c\approx$ 27 Hz  and $k_c\approx$ 10-11 cm$^{-1}$ as shown in figure \ref{capillary_waves}(\textit{f}). These frequencies show excellent agreement with the corresponding typical pressure oscillation frequencies measured in the vapor layer of star-shaped Leidenfrost ethanol and acetone drops (see figure \ref{scaling}\textit{c}). 

It is well-known that capillary waves can be generated at a liquid-vapor interface due to a strong shear stress in the vapor \citep{Miles1957,Zhang1995,Paquier2015,Zeisel2008,chang2002complex}. A similar mechanism underlies the Kelvin-Helmholtz instability, however, the intermediate Reynolds number in the Leidenfrost vapor layer complicates the analysis. Nevertheless, we can estimate the strength of this shear stress. Typically, the ``friction velocity" is generally used to measure the strength of shear, which is defined as $u_*=\sqrt{\tau/\rho_v}$, where $\tau$ is the shear stress at the liquid-vapor interface. The maximum shear stress at the interface is:
\begin{equation} 
\tau=\frac{6\eta_vu}{e},
\label{shear_stress}
\end{equation}
assuming a parabolic-flow profile in the vapor layer with mean velocity $u$ near the edge of the drop. Using water as an example, and plugging (\ref{u}) and (\ref{e}) and the value of $\eta_v$ of water vapor into (\ref{shear_stress}) yields $u_*\approx 2$ m/s. This friction velocity is quite strong and sufficient to lead to the growth of unstable modes with wavelengths of millimeter scale \citep{Zhang1995,Zeisel2008}.

The general dispersion relation of gravity-capillary waves with a dense upper-layer is: 
\begin{equation} 
f_c=\frac{1}{2\pi}\sqrt{\left(-gk_c+\frac{\gamma k_c^3}{\rho} \right)\tanh\left( k_ch \right)},
\label{capillary_wave_eq}
\end{equation}
where $k_c$ = 2$\pi$/$\lambda_c$, and $h\approx2 l_c$ is the thickness of the drop. For simplicity, we have assumed that the normal velocity is zero at the upper surface of the drop.  For a large Leidenfrost water drop whose characteristic pressure oscillation frequency in the vapor layer is $f_p\approx$ 26 Hz (see figure \ref{f_vs_R}), the corresponding capillary wavelength is calculated to be $\lambda_c\approx 3.03l_c$, so that $k_c\approx$ 8.3 cm$^{-1}$  \citep{Maleidenfrost2016}.  Similarly, for the Leidenfrost acetone and ethanol drops shown in figures \ref{capillary_waves}(\textit{a}), \ref{capillary_waves}(\textit{c}), and \ref{capillary_waves}(\textit{e}), the corresponding $k_c\approx 13 $ cm$^{-1}$, which is slightly larger than the positions of the peaks indicated in figures \ref{capillary_waves}(\textit{b}), \ref{capillary_waves}(\textit{d}), and \ref{capillary_waves}(\textit{f}), where $\lambda_c\approx 4l_c$. The agreement between the estimate for $k_c$ and the measurements shown in figure \ref{capillary_waves} is good considering the simplicity of (\ref{capillary_wave_eq}), which is derived using an inviscid, semi-infinite flow in both phases.

The capillary-wave origin for the star-shaped oscillations also agrees with the minimum size of the $n$ = 2 mode. More specifically, the radius we measured in our experiments is $R$ rather than $R_{\mathrm{neck}}$ as illustrated in figure \ref{sketch}. The relationship between these two lengths is $R=R_{\mathrm{neck}}+0.53l_c$ \citep{snoeijer2009maximum,burton2012geometry}. Thus, the minimum drop size required to fit one capillary wavelength beneath the drop is $2R_{\mathrm{neck}}\approx \lambda_c$. Using $\lambda_c\approx 3.03l_c$ from above, we find that the radius of an $n$ = 2 mode drop should be $R\approx2.05 l_c$. This is in good agreement with figure \ref{scaling}(\textit{a}), which shows $R/l_c$ is slightly less than 2 for the smallest drops. Taken together, this analysis suggests a purely hydrodynamic origin for the star oscillations based on capillary waves generated by a strong shear stress in the rapidly-flowing vapor beneath the drop.

\begin{figure}
\begin{center}
\includegraphics[width=4 in]{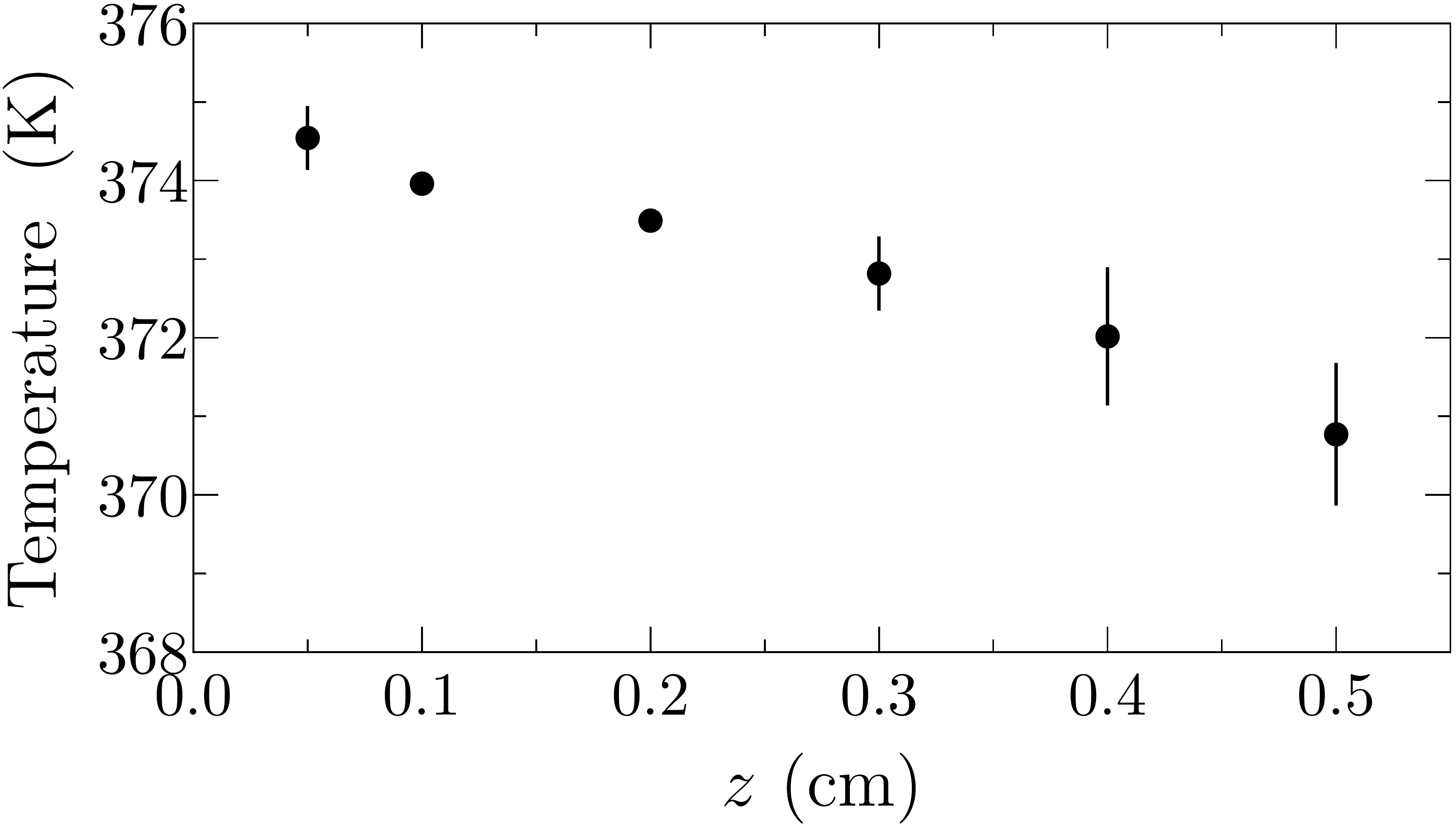}
\caption[]{(Color online) The temperature profile of a large Leidenfrost water drop with the volume of 10$^{-6}$ m$^{3}$. The error bars come from the standard deviations of multiple measurements. 
} 
\label{temperature_profile}
\end{center} 
\end{figure}

\subsection{Thermal effects}
\label{thermal effect}

In the Leidenfrost effect, thermal energy is transferred from the substrate to the drop, inducing evaporation and enabling the sustained star-shaped oscillations. Here we consider the role of thermal effects, such as convection in the liquid, which may play a role in the star oscillations. At sufficiently large temperature gradients in liquid layers, convective structures can develop with well-defined length scales. A well-known example is B\'{e}nard-Marangoni convection, where hexagonal patterns are generated in a thin layer of liquid when heated from the bottom \citep{benard1901tourbillons,rayleigh1916lix}. The convection pattern is strongly affected by the variation of surface tension with temperature \citep{marangoni1871ausbreitung,schatz2001experiments,maroto2007introductory}. Given that our data collapses when scaled by the capillary length of the liquid (figure \ref{scaling}), this particular type of convection may be important for Leidenfrost drops since the capillary length also appears in the length scale which characterizes the size of the convective patterns. 

\begin{figure}
\begin{center}
\includegraphics[width=5 in]{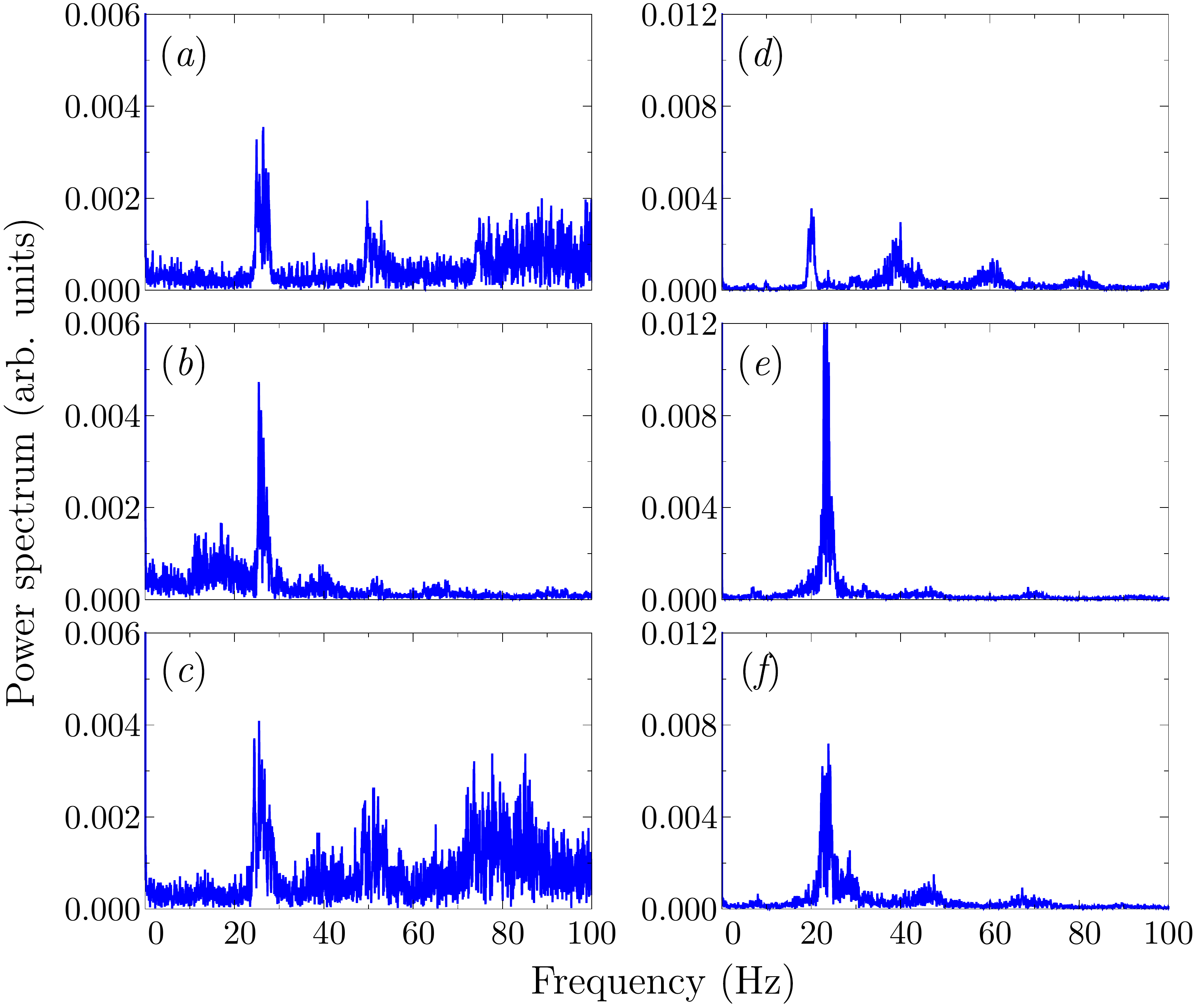}
\caption[]{(Color online) Fourier power spectra of the pressure variations in the vapor layer at different environment temperature $T_1$. (\textit{a}), (\textit{b}) and (\textit{c}) represent pressure variations of a 4-mode Leidenfrost water drop at $T_1 \approx$ 367 K, 483 K, and 623 K, respectively. (\textit{d}), (\textit{e}) and (\textit{f}) represent pressure variations of a 5-mode Leidenfrost water drop at $T_1 \approx$ 367 K, 483 K, and 623 K, respectively. All of the data shown here was obtained at the same scanning rate within the same time interval, 20 s.  
} 
\label{thermal_convection}
\end{center} 
\end{figure}

Generally, the Marangoni number, $Ma$, and Rayleigh number, $Ra$, are used to characterize B\'{e}nard-Marangoni convection. These dimensionless numbers are defined as: 
\begin{eqnarray}
\label{ma}
Ma=\frac{\left(\partial \gamma /\partial T\right)\delta Th}{\nu_l \rho D_l},\\
\label{ra}
Ra=\frac{g\alpha \delta T h^3}{\nu_l D_l}.
\end{eqnarray}  
These numbers characterize the strengths of thermocapillary and buoyancy effects, respectively. Additionally, the ratio of thermal conduction inside the liquid to the conduction at the liquid-vapor interface is characterized by the Biot number: 
\begin{equation}
Bi=\frac{\beta h}{\kappa_l}=\frac{T_{\mathrm{hot}}-T_{\mathrm{cold}}}{T_{\mathrm{cold}}-T_1},
\label{bi}
\end{equation}
The definitions of the symbols in $Ma$, $Ra$, and $Bi$ are listed in table \ref{tab:2}, as well as their values for water at the boiling point.

\begin{table}
\begin{center}
\def~{\hphantom{0}}
\begin{tabular}{llr}
     Symbol                                          &Quantity                                       &Value \\ 
     \hline
     $\nu_l$                                           &liquid kinematic viscosity	                      &2.94$\times$10$^{-7}$ m$^2$/s \\
     $D_l$	                                     &liquid thermal diffusivity 	                      &1.743$\times$10$^{-7}$ m$^2$/s   \\
     $\alpha$	                                     &thermal expansion coefficient          &7.52$\times$10$^{-4}$ /K   \\
     $\kappa_l$	                               &liquid thermal conductivity                       &3.2$\times$10$^{-2}$ W/m/K \\
     $h$	                                           &drop thickness	                            &0.005 m  \\
     $\delta T$	                              &temperature difference	                      &$T_{\mathrm{hot}}-T_{\mathrm{cold}}$ \\
     $\beta$	                                    &heat transfer coefficient	               &depends on $\delta T$  \\
     $\partial \gamma/\partial T$	    &surface tension gradient                  &1.46$\times$10$^{-4}$ N/m/K\\
     $T_1$	                                    &surface temperature	               &$>$367 K  \\
     \hline
\end{tabular}
\caption{Physical properties of water at the boiling points \citep{lemmon2011nist}.}
\label{tab:2}
\end{center}
\end{table}

Figure \ref{temperature_profile} shows the temperature profile of a large Leidenfrost water drop levitating on its vapor layer with $T_s= 673$ K, in which the substrate surface is defined as $z=0$. The temperature was measured with a fine-point thermocouple, as described in \S\ref{experiment}. We define the temperatures of the bottom and top surfaces of the drop as $T_{\mathrm{hot}}$ and $T_{\mathrm{cold}}$, respectively, and the temperature of the position which is slightly above the top surface of the drop is denoted as the environment temperature, $T_1$, as shown in figure \ref{sketch}. From figure \ref{temperature_profile} we can obtain $\delta T$ = $T_{\mathrm{hot}}-T_{\mathrm{cold}}\approx$ 3.8 K, the environment temperature measured to be $T_1=$ 367 K, thus we can calculate $Ma\sim 2\times 10^5$, $Ra\sim 7\times 10^4$, and $Bi\approx$ 1.1. Both the values of $Ma$ and $Ra$ are much larger than their critical values, which are typically of order 100 for the initiation of convective instability \citep{maroto2007introductory,leal2007advanced}. Thus it is possible that thermal convection plays a role in initiating the star-shaped oscillations. In this case, one may expect the star-shaped wavelength, $\lambda_d$, to be related to the critical wavelength of the convective instability. This critical wavelength depends on the Biot number. We implemented a qualitative test for this dependence by wrapping aluminum foil around the substrate and Leidenfrost drop, which dramatically increased the environment temperature, $T_1$, near the top of drop.  

Figures \ref{thermal_convection}(\textit{a}) and \ref{thermal_convection}(\textit{d}) show Fourier power spectra of the pressure oscillations in the vapor layer of Leidenfrost water drops during $n$ = 4 and $n$ = 5 oscillations at $T_1 \approx$ 367 K, respectively. Figures \ref{thermal_convection}(\textit{b}) and \ref{thermal_convection}(\textit{e}) are the spectra of pressure oscillations during $n$ = 4 and $n$ = 5 oscillations at $T_1 \approx $ 483 K, respectively. Finally, figures \ref{thermal_convection}(\textit{c}) and \ref{thermal_convection}(\textit{f}) represent the same modes at $T_1$ = 623 K. For $T_1$ = 483 K and 623 K, the direction of heat transfer at the upper surface of the drop has been reversed; energy is added to the drop. Overall, dramatically increasing the surrounding temperature affects the appearance of higher harmonics in the spectra. The behavior is non-monotonic, and is likely due to the highly nonequilibrium conditions (high evaporation rate) induced by the high temperatures of the substrate ($T_s$) and environment ($T_1$). A more detailed understanding of the pressure oscillation spectra is left for future studies. Nevertheless, the position of the main peak in the spectra is independent of $T_1$, indicating that convection and the details of thermal transport in Leidenfrost drops play a secondary role in the star-shaped oscillations.

\section{Conclusion and outlook}

Both large and small Leidenfrost drops display self-organized oscillations due to the constant input of thermal energy and continuous evaporation and flow beneath the drop. Here we have focused on radial oscillations (i.e. ``breathing" mode) of small Leidenfrost drops, and the large-amplitude, star-shaped oscillations that appear in large Leidenfrost drops. We have characterized the number of observed modes for various volatile liquids, the frequency and wavelength of the oscillations, and the pressure variations in the vapor layer beneath the drops. The number of observed modes is sensitive to the properties of the liquid (see table \ref{tab:1}), i.e. the star-shaped oscillations of smaller Leidenfrost drops are dissipated by the liquid viscosity, which sets the minimum oscillation mode number $n_{min}$ that can be observed in experiment. The relationship between the frequency and wavelength agrees very well the quasi-two-dimensional theory proposed by \citet{yoshiyasu1996self}. The dominant frequency associated with the pressure oscillations is approximately twice the drop oscillation frequency, consistent with a parametric forcing mechanism for the star oscillations. 

One of the main findings of our work is identifying the underlying cause of the pressure oscillations. By imaging the liquid-vapor interface from below the drop, and using a simplified model for the flow in the thin vapor layer, we conclude that capillary waves of a characteristic wavelength, $\lambda_c\approx 4 l_c$, lead to pressure oscillations at the experimentally measured frequency. The flow in the vapor layer is quite rapid. Near the edge of the drop, the mean radial velocity can reach 1-2 m/s or more. In the small gap ($\sim$ 100 $\mu$m) between the liquid and the solid surface, this flow applies a large shear stress to the liquid-vapor interface, and can easily excite capillary waves with millimeter-scale wavelengths. The dispersion relation for the capillary waves then leads to a characteristic frequency for the pressure oscillations, which in turn parametrically drive the star-shaped oscillations. Furthermore, although the vapor flow is inherently driven by evaporation and heat transfer, the substrate and surrounding temperature have little effect on the dominant frequency and wavelength of the oscillations, suggesting they are purely hydrodynamic in origin.

Although the work presented here has focused mostly on the origin of star-shaped oscillations, the coupling of the flow in the vapor layer and the liquid-vapor interface underlies a rich spectrum of dynamical phenomena observed in both Leidenfrost liquid layers and drops. In particular, of key interest is understanding the failure of the Leidenfrost vapor layer which can lead to explosive boiling. If the observed capillary waves beneath the drop act as the precipitant to vapor-layer failure, then it is possible that geometrical tailoring of the surface to be commensurate with $\lambda_c$ may inhibit the generation of capillary waves. In addition, patterned, ratchet-shaped substrates with with wavelengths $\approx$ 1-3 mm are known to induce propulsion of small Leidenfrost drops \citep{cousins2012ratchet,linke2006self,lagubeau2011leidenfrost}, however, less is known about transport of large quantities of liquid, and the dependence on the wavelength of the surface patterns. We leave these questions open to future experiments. More generally, our results may offer insight into the direct control of oscillations in levitated drops in many other systems \citep{haumesser2002high,duchemin2005static,paradis2005surface,ishikawa2006noncontact,lister2008shape,langstaff2013aerodynamic}, for example, precise control is crucial when levitating high-temperature or harmful liquids using a gas film. We also expect our results to enhance the understanding of dynamics that couple a thin, supporting gas film, a liquid interface, and a solid surface, a scenario which occurs through forced wetting and gas entrainment in liquid coating \citep{Xu2005,Driscoll2011,marchand2012,Kolinski2012,Liu2013,Liu2015}.

\section{Acknowledgements}
We thank Tom Caswell for providing the experimental data related to the breathing mode of Leidenfrost drops, and Juan-Jos\'{e} Li\'{e}tor-Santos for helpful discussions. This work was supported by the National Science Foundation through grant NSF DMR-1455086.

\bibliographystyle{jfm}
\bibliography{star}

\end{document}